\documentclass{aastex62}
\usepackage{amsmath,amssymb,amsfonts}
\usepackage{algorithmic}
\usepackage{textcomp}
\usepackage{booktabs}
\usepackage{bbm}
\usepackage{mathtools}

\newcommand{\Cov}{\mathrm{Cov}}

\DeclareMathOperator*{\argmin}{arg\,min}
\DeclareMathOperator*{\argmax}{arg\,max}
\newcommand{\LRT}[2]{%
    \mathrel{\mathop\gtrless\limits^{#1}_{#2}}%
}

\accepted{\today}
\submitjournal{AJ}
\shorttitle{A Search for Exoplanet Candidates in TESS 2-min Light Curves using Joint Bayesian Detection}
\shortauthors{Taaki, Kemball, \& Kamalabadi}

\begin{document}

\title{A Search for Exoplanet Candidates in TESS 2-min Light Curves using Joint Bayesian Detection}
\correspondingauthor{Jamila Taaki}
\email{xiaziyna@gmail.com}

\author[0000-0001-5475-1975]{Jamila S. Taaki}
\affiliation{Department of Electrical and Computer Engineering,
University of Illinois at Urbana-Champaign\\
306 N. Wright St. MC 702, Urbana, IL 61801-2918}

\author[0000-0001-6233-8347]{Athol J. Kemball}
\affiliation{Department of Astronomy, University of Illinois at Urbana-Champaign\\
1002 W. Green Street, Urbana, IL 61801-3074}
\affiliation{Wits Centre for Astrophysics, School of Physics, University of the Witwatersrand, P.O. Box Wits, Johannesburg 2050, South Africa}

\author{Farzad Kamalabadi}
\affiliation{Department of Electrical and Computer Engineering,
University of Illinois at Urbana-Champaign\\
306 N. Wright St. MC 702, Urbana, IL 61801-2918}
\nocollaboration
\begin{abstract}
In this work, we apply an exploratory joint Bayesian transit detector \citep{taaki2020bayesian}, previously evaluated using Kepler data, to the 2 min simple aperture photometry light curve data in the continuous viewing zone for the Transiting Exoplanet Survey Satellite (TESS) over three years of observation. The detector uses Bayesian priors, adaptively estimated, to model unknown systematic noise and stellar variability incorporated in a Neyman-Pearson likelihood ratio test for a candidate transit signal; a primary goal of the algorithm is to reduce overfitting. The detector was adapted to the TESS data and refined to improve outlier rejection and suppress false alarm detections in post-processing. The statistical performance of the detector was evaluated using transit injection tests, where the joint Bayesian detector achieves an 80.0 $\%$ detection rate and a $19.1 \%$ quasi-false alarm rate at a detection threshold $\tau = 10$; this is a marginal, although not statistically significant, improvement of $0.2 \%$ over a reference sequential detrending and detection algorithm. In addition, a full search of the input TESS data was performed to evaluate the recovery rate of known TESS objects of interest (TOI) and to perform an independent search for new exoplanet candidates. The joint detector has a $73 \%$ recall rate and a $63 \%$ detection rate for known TOI; the former considers a match against all detection statistics above threshold while the latter considers only the maximum detection statistic. 

\end{abstract}


\section{Introduction} \label{sec:intro}
\par{The Transiting Exoplanet Survey Satellite (TESS)\footnote{tess.mit.edu} \citep{ricker} was launched in 2018 to survey $\sim 10^6$ nearby stars for planets. TESS is the successor to two decades of wide-field exoplanet transit surveys: primarily CoRoT \footnote{sci.esa.int/corot} \citep{auvergne2009corot} and Kepler \footnote{keplerscience.arc.nasa.gov} \citep{borucki}. Such surveys are scientifically critical to understanding the population statistics of exoplanets by type and exoplanetary configuration \citep{zhu2021}. Innovations in statistical inference and specialized data processing have been critical in achieving the photometric precision \citep{Deeg2018} required to robustly detect exoplanet transits in these data-intensive surveys \citep{jara2020transiting}. In particular, a number of specialized approaches have been developed to remove residual instrumental systematics that may otherwise mask astrophysical transit signals in light curves \citep{twicken2010presearch, kep, Stumpe_2012, Stumpe_2014, roberts, aig2016, foreman}. In this work we adapt a prior algorithm for joint Bayesian transit detection and systematic noise characterization \citep{taaki2020bayesian} we previously evaluated on Kepler data, and apply it to TESS 2-min simple aperture photometry (SAP) light curves. The goals of this paper are to explore the modifications required and the numerical performance of this algorithm when applied to data with different residual instrumental systematics and to independently search for new candidate exoplanets in the TESS data.}
\par{
The TESS and Kepler missions differ significantly in their observing strategy and cadence. TESS has an instantaneous field-of-view (FOV) of $24^{\circ} \times 96^{\circ}$ spanned by four individual cameras each containing four $2048 \times 2048$ pixel CCDs in their focal planes \citep{ricker2016}. Individual pointings align the primary axis of the FOV along ecliptic longitude spanning a range of ecliptic latitudes from $\sim 6^{\circ}$ to the ecliptic pole \citep{ricker2016}. Each pointing of the FOV is of $\sim 27$d duration and is known as a TESS sector \citep{ricker2016}. Each observing year in the primary mission spans one ecliptic hemisphere over 13 sectors equispaced in ecliptic longitude and alternates to the other hemisphere the following year \citep{ricker2016}. Approximately 2$\%$ of targets are continuously observed for the full year in the overlapping polar region known as the Continuous Viewing Zone (CVZ) \citep{Barclay_2018, eisner}. TESS produces full-frame images (FFI) with a 30-min cadence which are analysed by the Quick-Look-Pipeline (QLP) \citep{Huang_2020}. The TESS Science Processing Operations Center (SPOC) pipeline \citep{spoc, jenkins16} also produces SAP light curves with 2-min cadence for a list of targets defined in the TESS Input Catalog (TIC\footnote{tess.mit.edu/science/tess-input-catalogue/}) \citep{stassun}.
To date, there have been $7525$ TESS Objects of Interest (TOI) identified, $2262$ of which are from the SPOC pipeline and a further $728$ have been promoted from Community TESS Objects of Interest (CTOI) to TOI. Currently, $620$ TOI have been confirmed as exoplanets\footnote{tess.mit.edu/toi-releases/}.

\par{In this work we apply our prior Bayesian detection method to the 2-min SAP light curves \citep{twicken:PA2010SPIE, jenkins16, morris:PA2020KDPH} obtained for the TESS CVZ targets over the first three years of observation. Each year of data is searched independently and where a target is observed in multiple years light curve data were not combined. A total of $5535$ year-long light curves were searched for exoplanets with orbital periods in the range $1 - 100$ days. Our search is complementary to the SPOC search for exoplanet candidates and other searches of TESS 2-min SAP data. These include the Planet-Hunters-TESS project \citep{eisner} which identified $90$ promising candidates in the first two years of TESS operations using a combination of visual inspection and semi-automated vetting. The Weird detector \citet{wheeler, chakraborty2020hundreds} was used to search the first year of TESS data for any interesting periodic phenomena using a technique based on phase dispersion minimization and report 28 planet-like signals. Further \citet{Wong} performed a phase curve study of 22 known transiting systems with light curves from the first year to obtain secondary-eclipse, brightness modulation measurements, as well as other transit properties. \citet{holdsworth} perform a search for, and identify 12, rapidly oscillating stars.  We note the implementation of the Systematics-Insensitive Periodogram \citep{Angus_2016, Hedges}, a Python tool to search TESS targets from the CVZ for long-period signals and identify stellar rotation rates. This method uses a noise model that incorporates systematics alongside sinusoid signal terms, which are jointly fitted to form a periodogram. 
TESS light curves have also been used to study stellar properties including an eclipsing binary (EB) catalog \footnote{http://tessebs.villanova.edu/} \citep{Prsa_2022}, stellar flare statistics \citep{Feinstein_2020} and asteroseismology \citep{lund2021tess}.}

Wide-field transit surveys retain unmodeled residual instrumental errors after initial calibration and processing that compromise their high target photometric sensitivity \citep{Deeg2018}. The instrumental origins are diverse \citep{2020ksci.rept2J, vanderspek2018} and the systematic noise in the light curves is generally correlated both temporally and spatially across the imaging sensor \citep{taaki2024}. The removal of residual systematics (detrending) is known to risk overfitting and distorting transit features \citep{Christiansen_2013, foreman}, particularly if the systematic and transit variability occur over similar timescales and are correlated. In addition, the residual systematics may mimic transit signatures, and consequently lower transit detection performance.
As mitigation, joint transit detection has been explored by \citet{foreman}, who derived non-Bayesian joint maximum likelihood estimates of transit signals and a low-rank linear systematic noise model. \citet{kov_joint} performed joint-fitting on K2 data but found no advantage over a sequential approach, citing false alarms that may occur due to underfitting of systematics. \citet{Garcia_2024} introduced a Python tool for computing joint likelihoods but replaced a white noise model \citep{foreman} with a time-correlated noise model. In the context of detrended TESS PDC-SAP flux light curves they found this method to improve the detection of transits in the presence of short-timescale stellar variability ($< 1$ day).
Our joint Bayesian detector \citep{taaki2020bayesian} is non-sequential and marginalizes over an empirical systematics prior and a stochastic noise prior, posing a binary hypothesis test for the light curve either containing a transit or not containing a transit. A likelihood ratio test universally maximizes \citep{Moulin_Veeravalli_2018} the probability of detection at any rate of false alarm for a binary hypothesis model. Our algorithm is summarized in Section~\ref{sec: det model} and full details, including the novelty and context of the algorithm relative to other methods, can be found in the original paper \citep{taaki2020bayesian}.

There are several major differences between the Kepler long-cadence data analyzed in our prior work and the TESS 2-min data, which necessitate modifications of our algorithm implementation. Relative to Kepler, a high fraction ($\sim 13 \%$) of $2$-minute TESS flux density measurements are missing or flagged. Each TESS observational sector comprises two $\sim$13.7 day orbits around the Earth and data collection ceases for several hours between each orbit to downlink the data collected during this time \citep{ricker2016}.  In addition to this periodic gap, data are flagged or missing due to a $2-6$ day momentum dump cycle \citep{vanderspek2018, toi_guerrero}. Although our algorithm supports non-uniform data and does not require gap filling or interpolation, the cadence of missing data presents ambiguity in the detected exoplanet period, in common with other TESS transit searches \citep{toi_guerrero}. In addition, each year-long SAP light curve contains $\sim 200,000$ data values and poses general computational challenges \citep{spoc}. For our algorithm, we performed detection tests on $10^8$ candidate transit models with a $10$-minute step size in candidate period and epoch. This is a high performance computing (HPC) challenge and was implemented in optimized form on the Blue Waters petascale system\footnote{https://bluewaters.ncsa.illinois.edu/hardware-summary}.

In addition to the cadence differences, TESS and Kepler have different instrumental systematics. TESS has a lower combined differential photometric precision (CDPP) \citep{Christiansen_2012, tess_photometric_noise}, achieving below 230 ppm for a 10th magnitude star over 1 hr \citep{fausnaugh2018tess}. Comparably, Kepler achieved 30 ppm for a 12th magnitude dwarf star over 6.5 hr (half the transit time of an Earth-analog) \citep{christiansen2010kepler, Gilliland_2011}. Consequently, the expected signal-to-noise ratio and detection power for transit signals is lower for TESS than would be expected in Kepler data. This lowered photometric precision arises from the highly elliptical 13.7 day TESS orbit. TESS data contain short-term instances of large flux variations \citep{fine_point}, arising from high pointing jitter, as well as quasi-periodic flux density ramps which are tied to the 2-6 day momentum dump cycles. During these cycles noisy reaction wheels amplify pointing jitter, blurring point spread functions (PSFs), and reduce apparent flux \citep{ricker_15}. After data downlink every 13.7 days, ramps occur due to thermal changes as a result of changing solar illumination during pointing. Additionally, scattered light features from the Earth and Moon exacerbate the reduction in apparent flux. In instances where the field is crowded, increased blending as a result of reduced per-pixel resolution may also contribute to lowered precision \citep{Lu_2020}. Reduced precision, as well as quasi-periodic systematics are known to affect transit event statistics and present as false positives in the TESS SPOC planet search \citep[p.17]{fausnaugh2018tess}, the TESS QLP search \citep{toi_guerrero}, and as described by \citet{Kunimoto_2023}. We adapted our algorithm implementation to filter false alarms using a robust metric similar to \citet{Seader_2013} in addition to several other post-processing metrics.

We evaluated our detector performance using single-transit injection tests on a sample of 1000 TESS CVZ light curves using limb-darkened transit signals. At a detection threshold of $\tau = 10$ the joint detector achieved a high detection rate of $\sim 80.0 \%$ with a $19.1 \%$ quasi-false-alarm rate. Our goal was to explore the joint detection algorithm on data with different systematics and we make no claim of optimality. The joint detector performs comparably to standard sequential detrending and detection, with a possible slight performance increment ($0.2 \%$). In our full search of CVZ light curves, approximately $28 \%$ of light curves have detections. We achieve a $63 \%$ detection rate for any TESS-objects-of-interest (TOI) present in CVZ light curves. The detection rate is defined as the rate of match between the TOI orbital period and the detected orbital period of the top statistic above the detection threshold, allowing for a 2-3 multiple of either period due to ambiguity in orbital periods described in \citep{toi_guerrero}. We achieve a $73 \%$ recall rate of TOI, where the recall rate is defined as the rate of match between the TOI orbital period and the best-matched orbital period above the detection threshold. After additional manual vetting to remove clear astrophysical false positives such as eclipsing binaries, as well as instrumental and statistical sources of false alarm we do not unambiguously identify any new candidate exoplanets.

\par The paper is organized as follows. Section \ref{sec:detmodel} describes the joint detector and implementation for TESS data. In Section \ref{sec:results} we report the results from our injection and detection results applied to TESS data. Section \ref{sec:discussion} contains a discussion of recovered candidates and conclusions are presented in Section \ref{sec:conclusion}. 

\section{METHODS} \label{sec:detmodel}

In this work, we apply a joint Bayesian transit detector \citep[{\it Detector A}]{taaki2020bayesian} to TESS CVZ SAP light curve data. In this section, we describe the refinements and implementation of this algorithm as applied to the TESS data.

\subsubsection{Detection Model} \label{sec: det model}

\par In this section we provide a concise summary of {\it Detector A} originally described by \citet{taaki2020bayesian}. We retain here the nomenclature of the original paper. After median subtraction and normalization, the light curve for star $i \in I$ (where $I$ is the  set of integers) is denoted as the vector $\mathbf{y}_i$ of length $N$ time samples. The light curve vector $\mathbf{y}_i$ is modelled as the vector sum of a systematic noise $\mathbf{l}_i$ term, a stochastic or statistical noise $\mathbf{s}_i$ term, and a possible transit signal $\mathbf{t} \in \mathbf{T}$ (where $\mathbf{T}$ is the parent set of transit signals). The term $\mathbf{s}_i$ includes both residual stellar variability, which is not modeled separately, and statistical error of a non-systematic nature. 

The transit detection framework is based on a binary hypothesis test: i) $H_0$: the null hypothesis that $\mathbf{y}_i$ contains no transit signal; and ii) $H_1$: the alternative hypothesis that $\mathbf{y}_i$ does indeed contain a transit signal $\mathbf{t}$ \citep[Equations (1) and (2)]{taaki2020bayesian}:
\begin{align}
H_0: \mathbf{y}_i &= \mathbf{s}_i + \mathbf{l}_i \\
H_1: \mathbf{y}_i &= \mathbf{t} + \mathbf{s}_i +  \mathbf{l}_i 
\label{eq:hypothesis}
\end{align}

The detection test can be stated as \citep[Equation (3)]{taaki2020bayesian}:
\begin{equation}
T({\mathbf{y}_i}) \equiv \mathcal{L}_{i}(\mathbf{y}_i) = \frac{p_1(\mathbf{y}_i)}{p_0(\mathbf{y}_i) } \LRT{H_1}{H_0} \tau
\label{eq:lrt}
\end{equation}
where $\mathcal{L}_{i}(\mathbf{y}_i)$ is a likelihood ratio test \citep{kaybook, wasserman_2013} which is Neyman-Pearson optimal, $p_h(\mathbf{y}_i)  \equiv p(\mathbf{y}_i | H_h) $ is the conditional probability of observing $\mathbf{y}_i$ under hypothesis $H_h \in \{H_0,H_1 \}$, and $\tau$ is the detection threshold.

\par This detector adopts a common low-rank linear model for the systematic noise contribution $\mathbf{l}_i$, expanded in terms of a set $\mathbf{v}_i$ of common basis vectors (CBV) across the sensor \citep{Kovacs_2005, Stumpe_2012}:
\begin{align}
    \bold{l}_i [n] = \sum_{k=1}^K  c^i_k \bold{v}_k [n]
    \label{eq: cbv}
\end{align} 
where $n$ denotes a discrete time sample index and $K$ is the model rank $(K \ll N)$ \citep[Equation (4)]{taaki2020bayesian}. Each $c_k^i$ is a coefficient weighting of $\bold{v}_k$ for light curve $i$. We denote the vector of coefficients for a light curve as $\bold{c}_i = [c_1^i, c_2^i, ... c_k^i]^T$ and the set of vector coefficients for all light curves on the sensor as $\{\bold{c}_i : i \in I \}$. 

\par The noise terms $\mathbf{s}_i$ and $\mathbf{l}_i$ are unknown a priori. This detector constrains the noise terms using empirical Gaussian priors $\mathbf{s}_i \thicksim \mathcal{N}(\mathbf{0}, \Cov_{\mathbf{s},i})$ and $\mathbf{c}_i \thicksim \mathcal{N}(\mu_{\mathbf{c}_i},\Cov_\mathbf{c})$, where the coefficients $\mathbf{c}_i$ act as proxies for the systematic noise $\mathbf{l}_i$ \citep{taaki2020bayesian}. The parameters of the models are obtained by empirical inference and, for the current application of the detector, this is described in Section \ref{sec:stoch_noise} and Section \ref{sec:sys_noise}. The joint noise is distributed as $\mathbf{l}_i + \mathbf{s}_i \sim \mathcal{N}(\mathbf{V} \mu_{\mathbf{c}_i}, \Cov_{s,i} + \mathbf{V} \Cov_{\mathbf{c}} \mathbf{V^T})$, where $V$ is an $N \times K$ matrix with $\mathbf{v}_k$ as column vectors \citep{taaki2020bayesian}. Using shifted data $\mathbf{\hat{y}}_i = \mathbf{y}_i - \mathbf{V}\mu_{\mathbf{c}_i}$ \citep{taaki2020bayesian} the detector may be equivalently formed for fixed signal $\mathbf{t}$ in the presence of joint noise $\mathbf{z}_i \thicksim \mathcal{N} (\mathbf{0} , \Cov_{\mathbf{z}, i})$, the joint noise covariance is $\Cov_{\mathbf{z}, i} = \Cov_{\mathbf{s},i} + \mathbf{V} \Cov_{\mathbf{c}} \mathbf{V^T}$.
When detecting a fixed signal in Gaussian noise, a matched filter is a Neyman-Pearson optimal detector with test statistic $T_i (\mathbf{\hat{y_i}})$ and equivalent to Equation~\ref{eq:lrt} \citep{taaki2020bayesian} and first used in Equation 1 of \citet{Jenkins_2002}:
\begin{align} \label{eq: detector}
T_i (\mathbf{\hat{y}}_i) = \frac{\mathbf{\hat{y}}_i^T \Cov_{\mathbf{z},i}^{-1} \mathbf{t}}{ \sqrt{\mathbf{t}^T {\Cov_{\mathbf{z},i}}^{-1} \mathbf{t}}} \LRT{H_1}{H_0} \tau
\end{align}

\par In this detector framework, exoplanet transits are searched for by computing transit detection tests $T_i (\mathbf{\hat{y_i}})$ for a discretized set of candidate signals $\mathbf{t} \in \mathbf{T}$. The functional form of candidate transit signals, as well as the discretization of the search space in the current application, are described in Section \ref{sec: transit_search}. 

\subsection{Implementation}\label{sec: implementation}
The implementation of the data processing pipeline that applies this detector to the TESS data is shown in Figure \ref{fig: pipeline_overview}. The flow of control moves from top to bottom and from left to right in the Figure. The multiple stages and tunable parameters for the pipeline are described in this Section. The nature of the TESS data requires implementation refinements relative to our previous application of the detector to Kepler data \citep{taaki2020bayesian}. These changes relate to: i) the observing cadence, flagged data characteristics, and multi-sector nature of the TESS data; ii) the computational optimization required to address the dataset size; and, iii) improved post-processing and filtering of candidate detections.

The implementation is described as follows. Section~\ref{sec: targets} describes the input light curve dataset and Section~\ref{sec: transit_search} defines the candidate transit search space $\mathbf{T}$. The nature of missing and flagged TESS data, and their treatment, is described in Section~\ref{sec: mask}. The choice of stochastic and systematic noise priors is described in Section~\ref{sec:stoch_noise} and Section~\ref{sec:sys_noise}. Post-detector processing is summarized in Section~\ref{sec: postprocess}. The computational complexity of the detector and its efficient multi-sector computation is described in Appendix \ref{sec: decompose}.

\subsubsection{Target Light Curves and SPOC CBV} \label{sec: targets}
As noted above, the input dataset to our detector pipeline comprises the TESS 2-min SAP light curves \citep{jenkins16} from the CVZ region for each of the first three years of operations. These SAP light curves were obtained from the output of the TESS SPOC pipeline, along with the single-scale and spike CBV provided by the SPOC analysis (see top left of Figure~\ref{fig: pipeline_overview}). The single-scale CBV are not separated in time-scale \citep{kep} in contradistinction to those formed using a multi-scale wavelet basis \citep{Stumpe_2014}; they are used here to form a light curve systematics prior and as an orthogonal systematics basis \citep{Stumpe_2012, kep}. The spike CBV contain large-amplitude narrow temporal spikes \citep{Tenenbaum2018TESSSD,lund2021tess}; they are used here during detection filtering and vetting.

The CVZ has a high degree of data completeness and allows the detection of long-period signals. In years 1 and 3 TESS observed in the southern ecliptic hemisphere and in year 2 in the northern ecliptic hemisphere. The number of targets per year is 1851 (year 1), 2046 (year 2), and 1922 (year 3). The total number of unique observed targets in our dataset is 5535. We note that the TESS 2-min cadence targets varied between the primary (years 1 and 2) and extended mission (year 3) \citep{Fausnaugh_2021}.

\begin{figure}[htbp]
  \centering
  \includegraphics[width=500pt]{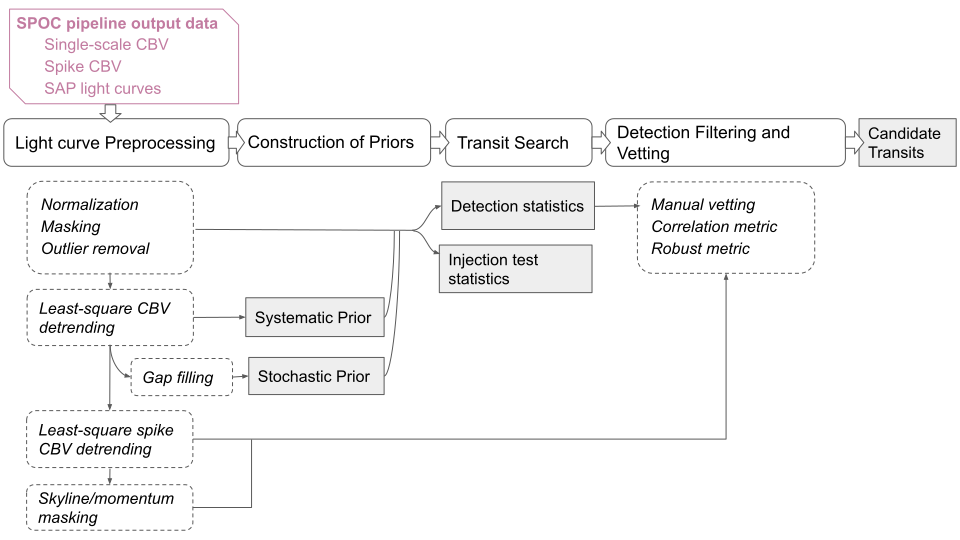}
  \caption{A schematic of the detector data processing pipeline. The input dataset obtained from the SPOC pipeline output is shown in pink. Processing steps are shown in rounded boxes: major processing steps are shown with a solid outline with constituent component processing steps below with dashed outlines. Boxes indicating data output products are shaded in grey.} 
\label{fig: pipeline_overview}

\end{figure}

\subsubsection{Normalization, Masking, and Outlier Removal} \label{sec: mask}

As the first step in preprocessing (Figure~\ref{fig: pipeline_overview}) each input light curve $\bold{y}_{in}$ was median normalized as $\bold{y}=\bold{y}_{in} / {\rm med} (\bold{y}_{in})-1$, where ${\rm med} (\bold{y}_{in})$ is the median of $\bold{y}_{in}$. Non-numeric (NaN) values were excluded in the normalization.

As noted above, TESS has a high rate of missing ($\sim 9\%$) and flagged ($\sim 4\%$) data compared to Kepler and at a significantly different cadence. Missing data arise due to a 2-6 day momentum dump cycle, a $\sim 13.7$ day downlink cycle, and stray light from the Earth and Moon \citep{vanderspek2018, toi_guerrero}. Figure \ref{fig:miss3} shows the average density of missing data in the year 1 input dataset light curves over observing sector and time. Figure \ref{fig:miss2} shows the breakdown of quality flagged data by systematic effect over year 1. We masked all missing and quality-flagged cadences \citep{jenkins16, vanderspek2018}. 
Outliers were further removed in each light curve while retaining negative values by using $3\sigma$ positive iterative thresholding:
\begin{equation}
\mathbf{y} [n] = 
\begin{cases}
     \mathbf{y}_{in}[n],& \text{if } \mathbf{y}_{in}[n] \leq 3 \sigma' \\
     \mathbf{y}[n] \sim \mathcal{N}(0, \sigma'),              & \mathbf{y}_{in}[n] > 3 \sigma'
\end{cases} \label{eq:3sigma}
\end{equation}
$\sigma' = \sigma (\mathbf{y}_{in}|_{|\bold{y}_{in}| \leq 3\sigma (\bold{y}_{in})})$, where $\sigma(\mathbf{y})$ is the sample standard deviation of $\mathbf{y}$ and $\mathbf{y}|_{f(\bold{y})}$ denotes a sub-vector of $\bold{y}$ where the condition $f(\mathbf{y})$ is true. In our prior work searching Kepler data \citep{taaki2020bayesian} we did not perform outlier removal or data masking; these algorithm refinements were added here to improve performance further.

\begin{figure}[ht!]
\centering
\includegraphics[width=.5\linewidth]{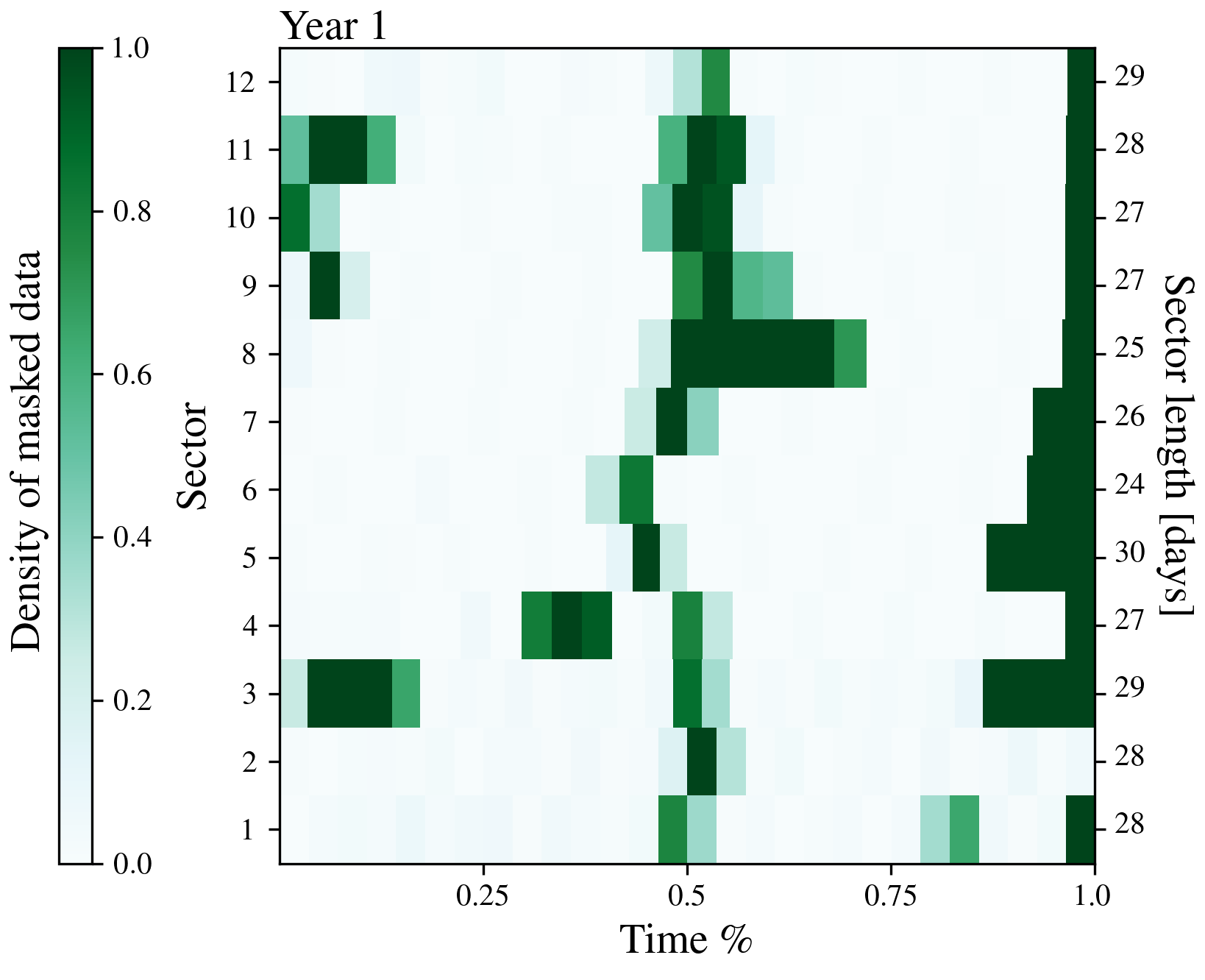}
\caption{The fraction of missing data in the year 1 light curves, averaged per day and across all target light curves. Each row represents a sector of data. Sectors are variable length, the total number of days in a sector is displayed on the right ticks. Each block represents the average fraction of data excluded (due to a quality flag or missing cadence) among the selected light curves, averaged over a day (approximately 720 cadences). Over each sector, most missing data occurs at the mid-point due to a data downlink. However, missing data also occurs throughout the remaining observational window. }
\label{fig:miss3}
\end{figure}

\begin{figure}[ht!]
\centering
\includegraphics[width=.5\linewidth]{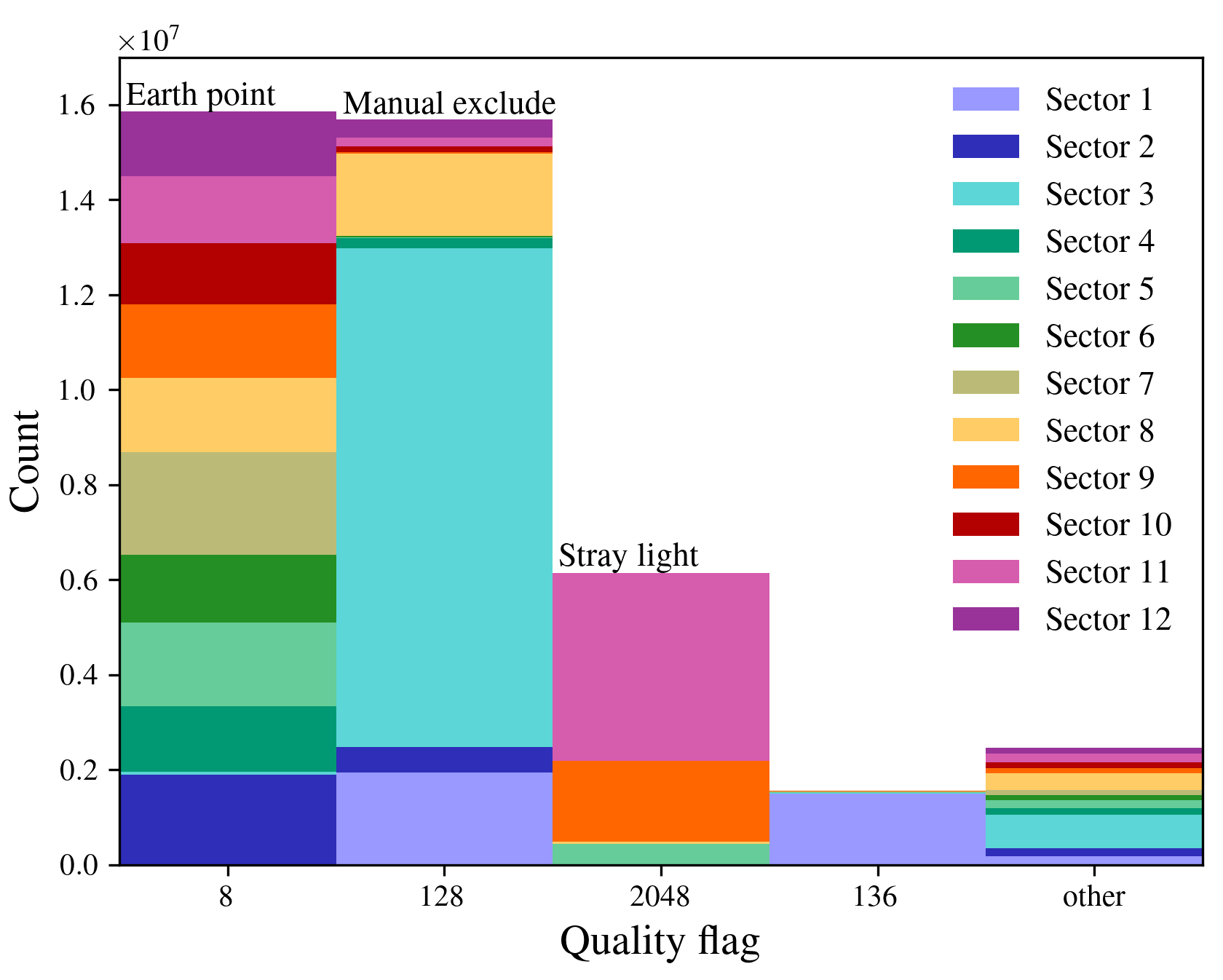}
\caption{The frequency of quality flagged cadences as they occur in year 1 light curves. The most commonly occurring quality flags have the following keys and descriptors: 8) spacecraft is in Earth point; 128) manual exclude due to an anomaly; 2048) stray light from Earth to Moon in camera FOV; 136) unspecified; and other) all other quality flags. Broadly, the majority of flagged data is due to the spacecraft being in Earth point, and uniformly occurs over different sectors. We justify masking all flagged data as the three most commonly occurring flags are dominant and indicate a high likelihood of photometric error. }
\label{fig:miss2}
\end{figure}

\subsubsection{Detrending and Construction of Coefficient Systematic Prior} \label{sec:sys_noise}

In earlier work \citep{taaki2020bayesian} the basis set $\{\mathbf{v}_k: k \in K \}$ was estimated using principal component analysis (PCA) on the collection of light curves. In this implementation, the TESS single-scale cotrending basis vectors (CBV) were adopted directly as the basis set $\{ \mathbf{v}_k : k \in K \}$ because the TESS CBV are based on a similar technique to PCA, the Singular Value Decomposition (SVD), and have also undergone additional data-filtering and analysis to minimize overfitting non-systematic variability \citep{kep, Stumpe_2012}. TESS provides an independent set of 8 or 16 such CBVs for each sector, camera and CCD set. An example set is shown in Figure \ref{fig:cbv}. We note that systematic flux ramps occurring before and after data downlinks (midway at 13.7 days and at the end of a sector) are removed in the TESS analysis before the CBVs are formed. Accordingly, these flux ramps are not removed in our detrending analysis, the effect of which is described in further detail in Section \ref{sec:results}.  

A least-squares fit of the basis to a light curve (Equation~\ref{eq: cbv}) without prior constraint may overfit and distort transits \citep{kep, Stumpe_2012}. As in \citet{taaki2020bayesian} we formed a systematic prior $p(\mathbf{c}) \sim \mathcal{N}(\mu_{\mathbf{c}_i}, \Cov_{\mathbf{c}_i})$ on the coefficients $\mathbf{c}_i$ that is marginalized over in the detector and constrains overfitting. We formed a model of the covariance of coefficient vectors $\Cov_{\mathbf{c}}$ from the sample covariance from least-square fits to all preprocessed light curves (including those outside the CVZ) for each set of sector, camera, and CCD combination. As in \citet{taaki2020bayesian}, although each least-square fit may be susceptible to overfitting, assuming independent astrophysical signals leads to a reasonable assumption of unbiased sample statistics.

The mean coefficient vector $\mu_{\mathbf{c}_i}$ was obtained from the least squares fit for the light curve under consideration. In Figure \ref{fig:samplefit} coefficient priors are shown alongside a histogram of fitted least-square fits from all light curves among a chosen sector CCD and channel. A Gaussian prior is used here as a reasonable choice for the broad light curve sample, as it inherently regularizes the data and easily captures correlations between systematic coefficients \citep{taaki2020bayesian}. However, empirical priors modeled on physical parameters as in PDC-MAP \citep{twicken2010presearch} may be more realistic and thus the Gaussian prior may be underpowered over a subset of light-curves. An empirical prior is not tractable for our Bayesian estimator.

\begin{figure}[ht!]
\centering
\includegraphics[width=.5\linewidth]{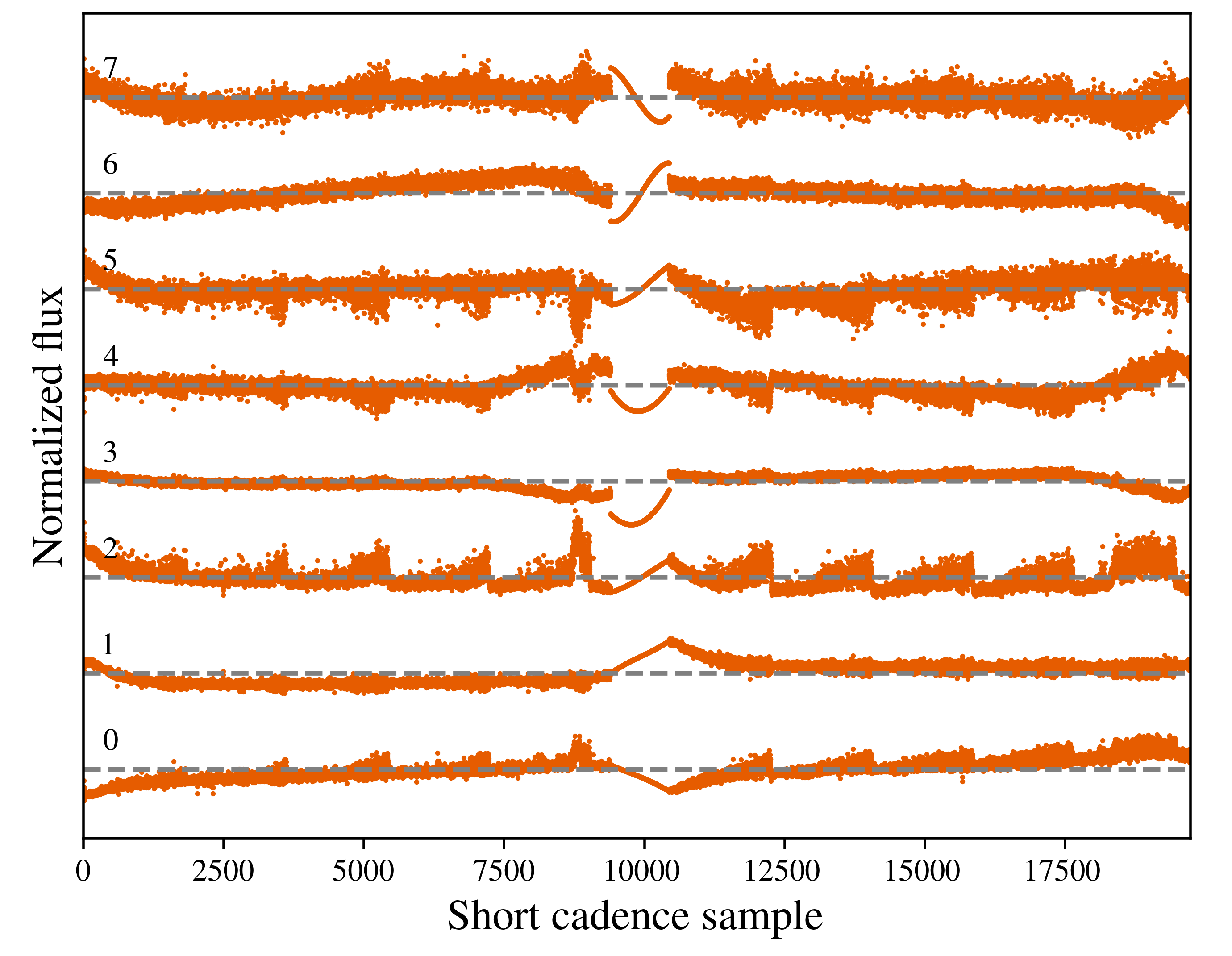}
\caption{An example of $8/16$ TESS cotrending basis vectors (CBV) for sector 2, camera 4, and CCD 3 are shown at a 2 min cadence. The CBVs capture systematics induced by the TESS sensor, in particular, the $2-6$ d visible periodic pointing jitter. Ramps at the beginning of each 13.7 d orbit are likely due to thermal effects following the pointing slew during sector downlink. The sector downlink at $\sim 13.7$ d appears in the middle of the plot. The data in this gap are masked in the current analysis.}
\label{fig:cbv}
\end{figure}

\begin{figure}[ht!]
\centering
\includegraphics[width =\linewidth]{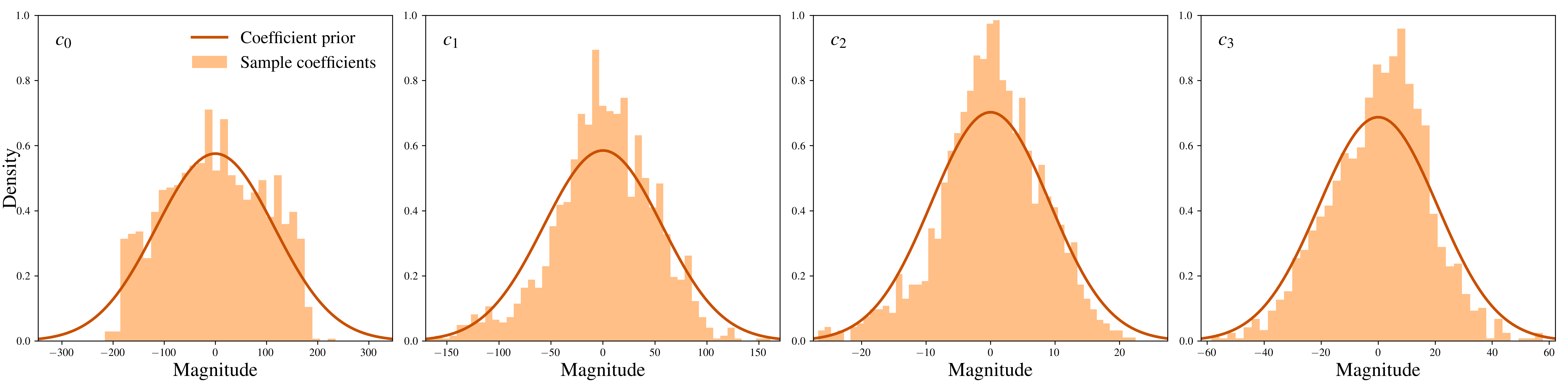}
\caption{The density of empirical least-squares coefficients $\hat{\mathbf{c}}$ for the four leading basis vectors in Sector 33, camera 4 and CCD 4, chosen as a random representative example. A fitted Gaussian prior is overlaid. The samples are heavy tailed in some cases, however, reasonably close to the fitted Gaussian. }
\label{fig:samplefit}
\end{figure}

\subsubsection{Construction of the Stochastic Prior} \label{sec:stoch_noise}
The stochastic noise includes primary contributions from stellar variability, read noise, and Poisson noise \citep{Gilliland_2011} which can be reasonably assumed to be asymptotically Gaussian $\bold{s} \sim \mathcal{N}(0, \Cov_{s})$ in combination. As in our prior work \citep{taaki2020bayesian}, we assumed the stochastic noise is zero-mean, stationary (per sector), and completely described by the covariance $\Cov_\mathbf{s} \in \mathcal{R}^{N \times N}$. As in that work, we estimate $\Cov_{\mathbf{s}, i}$ for light curve $i \in I$ in each sector using the spectrally smoothed autocorrelation computed from least-square detrended and $3\sigma$ thresholded light curve data. Computing the estimated and smoothed spectrum requires uniform time sampling, gaps in the TESS light curves were filled with randomly selected portions of the light curve from the same sector. However, as shown in Figure~\ref{fig: pipeline_overview}, gap filling is only required for this spectral estimation of the stochastic prior; it is not required for computing the detection statistic (Equation~\ref{eq: detector}). 

Estimates of $\Cov_{\mathbf{s}, i}[t,u] : |t-u| = k$ for larger lags have fewer data samples and therefore greater estimation variance. 
The unsmoothed sample spectrum values, corresponding to the Fourier transform of the light curve samples squared, are each independently $\chi^2_2$ distributed in the white noise case \citep{adaptive_kay, VanderPlas_2018}. The frequency binning of the spectral values averages independent $\chi^2_2$ variables reducing the variance of spectral values. The estimation variance of autocorrelation samples, as the inverse Fourier transform of the smoothed periodogram samples, is thus reduced. This has the effect of suppressing the covariance at large lag values.
We use a spectral smoothing window of $3$ frequency bins. In Figure \ref{fig: acv} an example of a light curve and the associated spectrally smoothed autocorrelation estimate is shown and compared to the sample autocorrelation, which has been suppressed at larger lags.

We note that this estimator is non-parametric and empirical. The spectral smoothing regularizes the stellar covariance model but may not entirely remove correlations due to transit eclipses. Physically motivated covariance models for stellar activity are described by \citet{angus} and \citet{rasmussen2003gaussian}.

Figure \ref{fig:std_flux} shows a sample of light curve noise models over sector 14 of the year 2 data. The relative 1$-\sigma$ contribution of the stochastic and systematic noise terms are shown over time. The noise models are highly dependent on the light curve under consideration.

\begin{figure}[ht!]
\centering
\includegraphics[width=.5\linewidth]{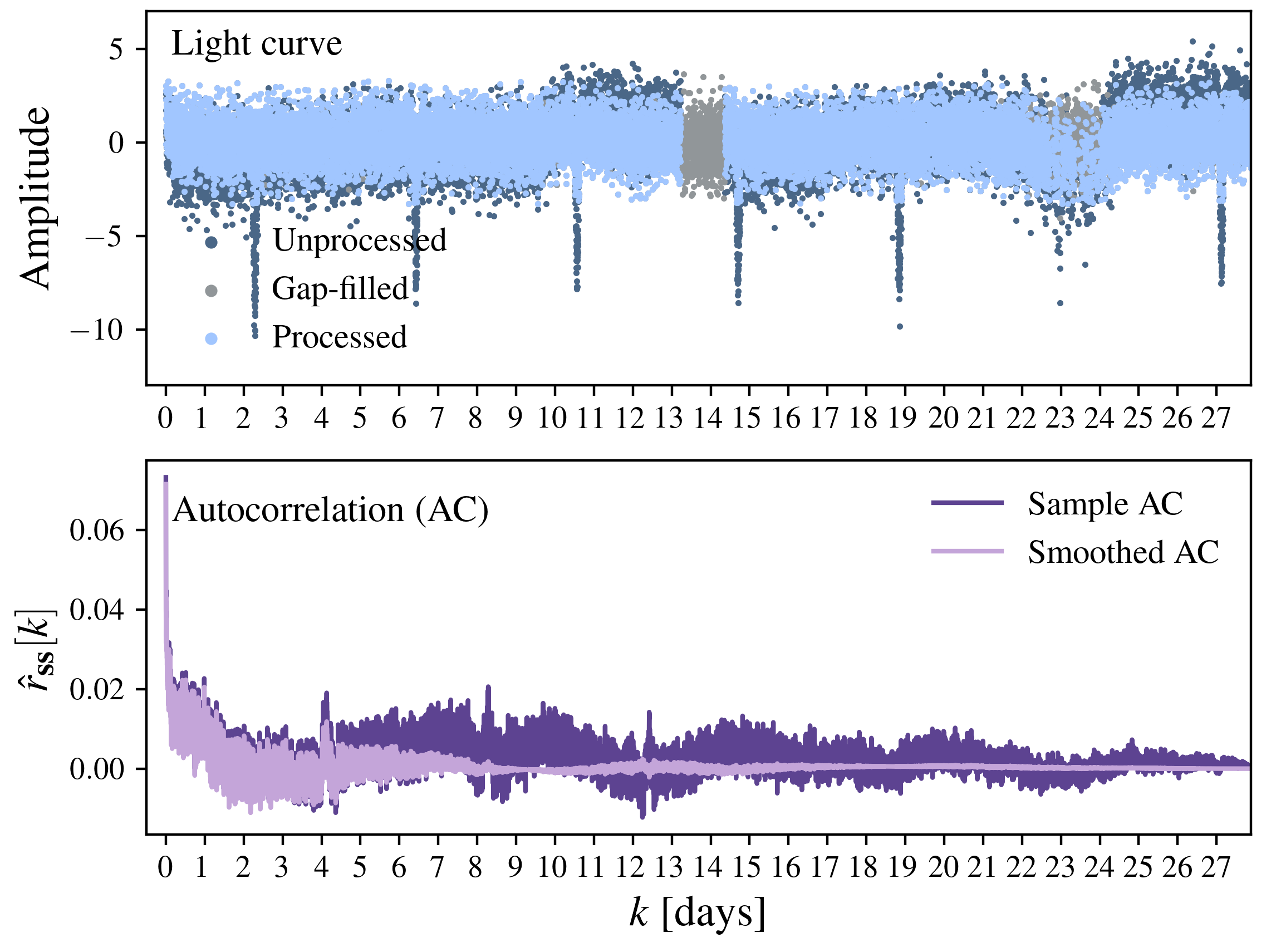}
\caption{ An example light curve (TID: 349518145, sector 1) with a transit signal is shown (top) and the associated fitted autocovariance $\Cov_{\mathbf{s}, i}[s-t] = f(|s -t|) = f(k) =  \hat{r}_{ss}(k)$ (below). In the top panel, the unprocessed light curve is shown in dark blue (background); this light curve is detrended and thresholded shown in light blue, and gap-filled with random samples shown in gray. The resulting light curve of light blue and gray points (foreground) is used to estimate the stochastic sample autocovariance (below, dark purple). The smoothed periodogram is shown in light purple. The smoothing reduces estimator variance and suppresses unfiltered transit variability; however, it may also suppress stochastic variability.}
\label{fig: acv}
\end{figure}

\subsubsection{Transit Search: Detection Tests and Transit Search Space} \label{sec: transit_search}
A transit search was performed on each light curve $i \in \mathcal{I}$ to find any transit signal $\mathbf{t} \in \mathbf{T}$ with a detection statistic (Equation~\ref{eq: detector}) above the threshold $T_i (\mathbf{\hat{y}}_i) > \tau$. As in \citet{taaki2020bayesian} we adopted a search space $\mathbf{T}$ of normalized box transit functions $\mathbf{b}$ parametrized by transit duration $d$, period $P$, and transit epoch $e$:

\begin{equation}
\mathbf{b}_{P, e, d} [n] = 
\begin{cases}
     1 ,& \text{if } (n-e)\mod_P \leq d \\
    0,              & \text{otherwise}
\end{cases}
\end{equation}

The detector is invariant to the transit depth as described in Appendix \ref{sec: decompose} and this parameter was therefore not included in the search space.

As shown in Appendix \ref{ap : box_approx}, for our matched filter detector, the reduction in detection power using a box transit function of duration $d'$ instead of a true limb-darkened transit of duration $d$ \citep{mandel2002analytic} is given by $ \frac{\mathbb{E}[T(\mathbf{y} ; \mathbf{b}_d)]}{\mathbb{E}[T(\mathbf{y} ; \mathbf{t}_d)]} = \mathbf{t}_d^T\mathbf{b}_{d'}$.  When $d' = d$, we found in numerical simulation that $\frac{\mathbb{E}[T(\mathbf{y} ; \mathbf{b}_d)]}{\mathbb{E}[T(\mathbf{y} ; \mathbf{t}_d)]} \approx 0.92$. However, the optimal box that produces the minimal reduction in detection power may be obtained by a slightly narrower duration $d' < d$ in practice. Allowing $d'$ to vary among the discrete set of durations searched in Table \ref{tab:search_space} and simulating numerically with 1000 limb-darkened single-transit events as described in Section \ref{sec: sim}, we found $\frac{\mathbb{E}[T(\mathbf{y} ; \mathbf{b}_d)]}{\mathbb{E}[T(\mathbf{y} ; \mathbf{t}_d)]} \approx 0.96$. 

We excluded all candidate transits with fewer than three transits present in the data, and required each transit to occur on a minimum of one non-masked cadence. For each year of data, we evaluated approximately $|\bold{T}| \sim 2 \times 10^8$ candidate transit signals per light curve, where the transit search space $\mathbf{T}$ was parameterized as in Table~\ref{tab:search_space}.  The step sizes in duration $d$, period $P$, and, epoch $e$, were chosen to afford a high level of correlation between a real transit signal and the best match candidate signal \citep{1996Icar, jenkins_2010}. As candidate durations under 1 hr are possible \citep{Stassun_2019}, and as further informed by the range of the Kepler Transiting Planet Search \citep{Kepler_TPS}, our transit duration range spans 1-16 h. 

A slight mismatch in period $P$ may significantly affect detection statistics; a full exploration of template mismatch is described by \citet{ 1996Icar, Seader_2013}. The maximum error in period $P$ is half the period step size: $\frac{\delta P}{2}$. After $M$ single transit events, the mismatch in the position of the $M^{th}$ transit event will be $\frac{M \delta P}{2}$.
Assuming a period $P$ with this worst case period mismatch, the epoch $e$ which would minimize the worst-case single transit offset and therefore provide the closest signal match in a search, is that which provides the best match for (approximately) the middle transit event. In this case, the maximum offset of a single event is approximately the number of transits in half of the light curve $\frac{N}{2P}$ multiplied by the maximum error in period as: $ \frac{N}{2P} \cdot \frac{\delta P}{2} = \frac{N \delta P}{4P}$. Since mismatch error is greater for shorter periods, we use a finer step size of $\delta P = 10$ min for periods $P \leq 27$ d and $\delta P = 20$ min for $P > 27$ d.
As an example, a transit signal with period $P = 8$ days can incur an approximate worst-case error in the location of a single transit event of $\sim 2 $ h; if the transit duration is less than $2$ h the template and transit event may be entirely mismatched. When calculating the matched filter test (Equation \ref{eq: detector}), each candidate transit in a sector was allowed to shift $\pm 2 $ h in epoch when finding the maximum detection statistic within this window. Transit events were not allowed, however, to move relative to one another in a sector.  This approach is also more robust to transit timing variations. In prior work \citep{taaki2020bayesian} we used the phase correlation method to find the best matched epoch $e$ of the template light curve. Here we do not utilize this method due to the phase-correlation method requiring multi-sector stitching and gap-filling of data. 

\begin{deluxetable}{ccc}[ht!]
\tablecaption{Transit Search Space \label{tab:search_space}}
\tablehead{
\colhead{Transit Parameter} & \colhead{Range} & \colhead{Step Size}}
\startdata
Orbital period  & $P \in [1, 100]$ d & $\delta P \in \{10, 20\}$ min \\
Transit duration & $d \in \{1,2,3,4,6,8,10,12,14,16 \}$ h &    \\
Epoch  & $e \in $ [0, $P$] & $\delta e \in \{10, 20\}$ min
\enddata
\tablecomments{The sample integration time is $\triangle t_{LC}=2 $ min. The step size of period $\delta P$ and epoch $\delta e$ is $10$ min for candidate periods less than the length of a sector $\sim 27 $ days, otherwise $\delta P$ and $\delta e$ are $20$ min.}
\end{deluxetable}

\subsubsection{Post-processing} \label{sec: postprocess}

In this application of our algorithm \citep{taaki2020bayesian} to TESS data we have improved robustness against false alarm detections. The detector form (Equation~\ref{eq: detector}) may produce spuriously high detection statistics \citep{Twicken_2016, Seader_2013} in the presence of light curve outliers, non-Gaussian noise, unmodeled residual systematics, or astrophysical mimics of exoplanet transits including eclipsing binaries and stellar harmonics. For TESS, unmodeled residual systematics are significant across the data gaps at 13.7 d and 2-6 d especially due to the data downlink and momentum dump cycle respectively. In this Section we describe post-processing and filtering used to reduce false alarms.

\begin{figure}[ht!]
\centering
\includegraphics[width=\linewidth]{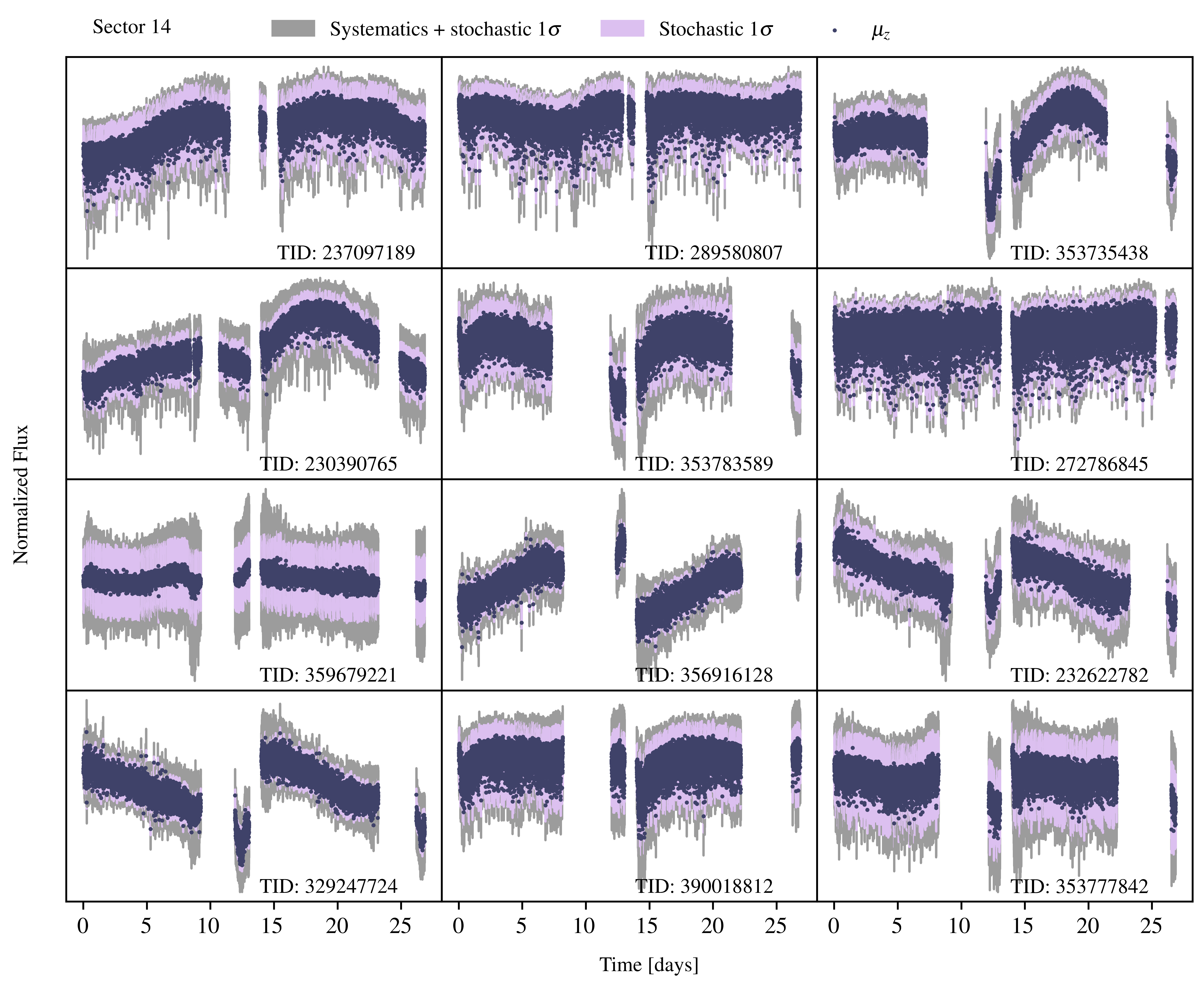}
\caption{The $1\sigma$ stochastic and systematic noise levels over time are shown for a random selection of light curves from year 2, sector 14. The mean noise term $\mu_{\mathbf{z}}$ (dark purple), the $1\sigma$ stochastic noise level $\sqrt{\Cov_{\mathbf{s}}[n,n]}$ (light purple), and the $1\sigma$ joint stochastic and systemic noise level $\sqrt{\Cov_{\mathbf{z}}[n,n]}$ (gray) are overplotted for each light curve. These noise covariance terms are described in Section \ref{sec: det model}. Although the relative contribution of the stochastic and systematic noise varies, they are generally of a similar magnitude. The masking is highly non-uniform between light curves in the same sector. We note that off-diagonal covariance terms are not depicted here; the full covariance of the light curve in the first panel (TID: 237097189) is shown in Figure \ref{fig: sample_cov}.  }
\label{fig:std_flux}
\end{figure}

\begin{figure}[ht!]
\centering
\includegraphics[width=.5\linewidth]{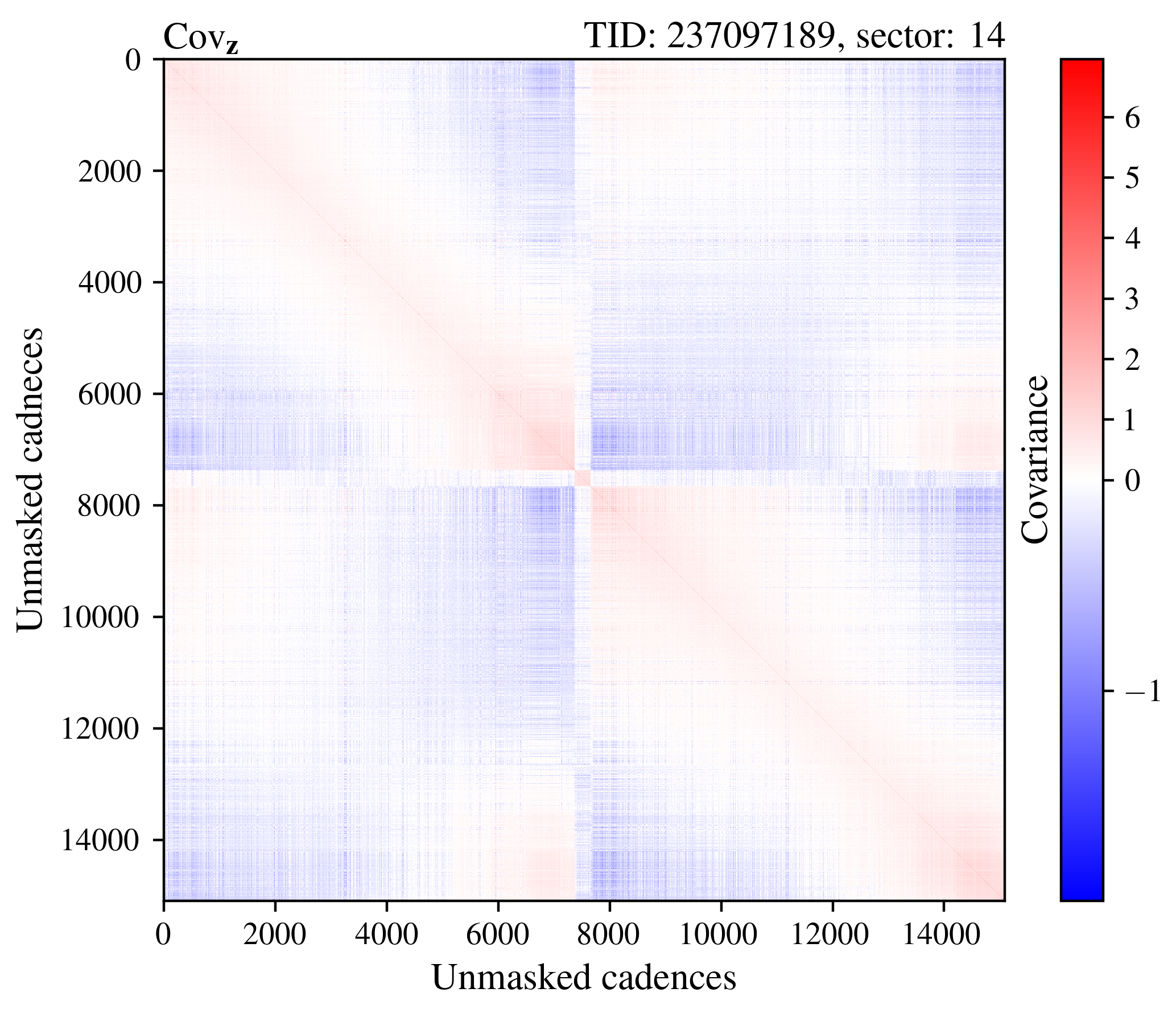}
\caption{ A sample joint-covariance matrix is shown for a light curve. The noise exhibits a diagonal component of positive correlation due to short-term correlated noise described by $\mathbf{s}$, as well as systematic variability described by $\mathbf{l}$. Longer term correlations are also seen with a 13.7 day blocked structure due to the dominant systematic, the 13.7 day downlink cycle. The dominant systematic is driven by the 13.7-day period orbit of TESS about the Earth. During the orbit the spacecraft experiences thermal changes and varying degrees of scattered light from the Earth and the Moon. }
\label{fig: sample_cov}
\end{figure}

\paragraph{Correlation Statistic} \label{sec: filter}
As a first step we computed the correlation coefficient $r_{\mathbf{ty}'}=\frac{\mathbf{t}^T \mathbf{y'}_i}{\|\mathbf{t} \|\| \mathbf{y'}_i\|}$ between the identified transit signal $\mathbf{t}$ and the associated processed light curve $\mathbf{y'}_i$. These light curves were processed using least-square detrending against the SPOC CBV (Section~\ref{sec:sys_noise}) but here also including the SPOC spike vectors \citep{jenkins16}. In addition, further masking was applied to these processed light curves $\mathbf{y'}_i$ beyond that described in Section~\ref{sec: mask}. Specifically, cadences within 1 h of momentum dump times were masked. In addition, a skyline histogram\footnote{github.com/christopherburke/TESS-ExoClass} (Figure~\ref{fig: skyline}) was used to identify individual cadences contributing disproportionately to transit event detections in multiple light curves and therefore likely to contain unmodeled residual systematics. As shown in red in this Figure, approximately 1000-3000 cadences were masked for each year of data because of their anomalous contribution to detection statistics. Figure~\ref{fig:skyline_plot} shows the mean (in blue) of the detrended and masked light curves for all detections in the period range [1,20] d in the sector 1 data overlaid with a detection density plot per cadence (gray); points filtered by the skyline histogram analysis are shown in red.

The correlation coefficient was computed within a window tolerance about the expected location of a transit event as described in Section \ref{sec: transit_search}. All detections with a correlation metric $r_{\mathbf{ty}'} < 0.1$ were filtered out as potential false alarms. This removed approximately one third of initial detections. 

\begin{figure}[ht!]
\centering
\includegraphics[width=.5\linewidth]{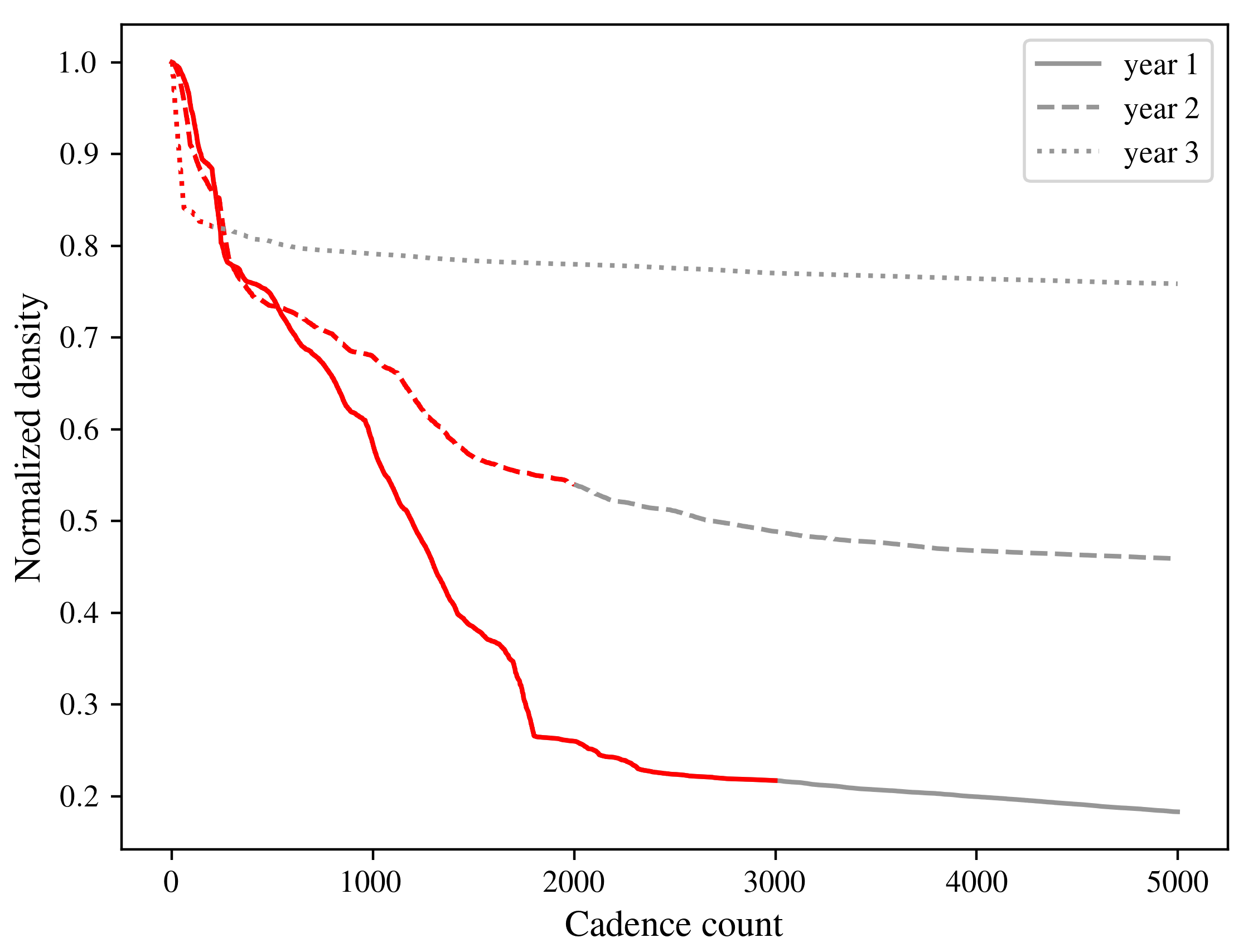}
\caption{Sorted skyline histogram of the relative contribution of individual cadences to the detection statistics (as a normalized density) for all light curves. The x-axis markers indicate the number of individual cadences to the left of each marker, sorted in order of transit detection contribution (density). Separate histograms are plotted for the year 1, 2, and 3 data. The sort order of individual cadences by density differs by year and the x-axis is cadence count. The red part of the curve indicates the knee cutoff used to mask cadences with anomolous density, and therefore likely unmodeled residual systematic errors. More cadences are filtered in the year 1 and 2 data compared to year 3.}
\label{fig: skyline}
\end{figure}

\begin{figure}[ht!]
\centering
\includegraphics[width=.5\linewidth]{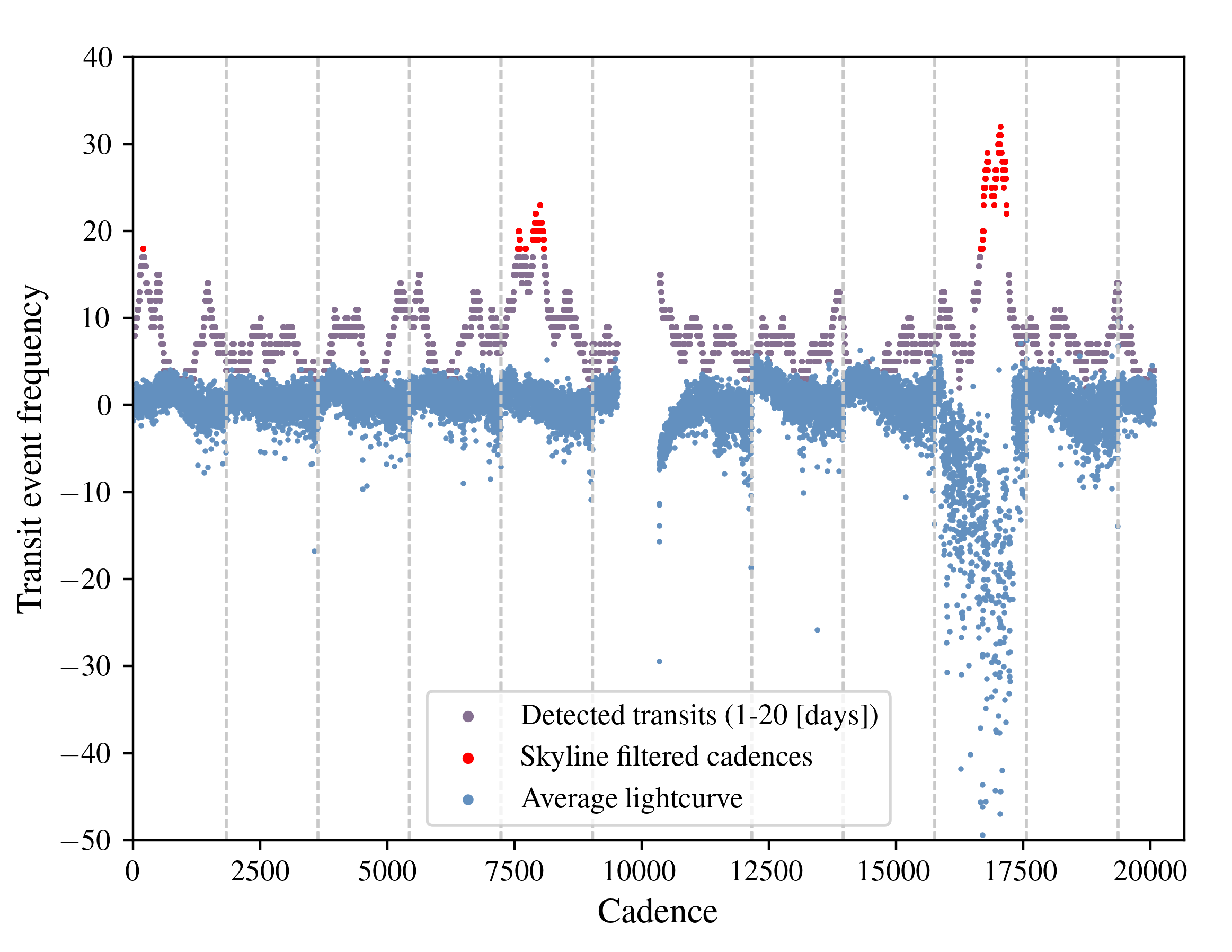}
\caption{A skyline plot of the mean light curve (blue) here formed as the average over all detrended and masked detection light curves with a period in the range [1,20] d in the sector 1 data. The vertical dashed gray lines indicate momentum dump events. The purple light curve is proportional to the frequency with which each individual cadence appears in any detected transit. Those cadences marked in red were filtered based on a skyline histogram analysis (Figure~\ref{fig: skyline}). Significant uncorrected systematics are visible in the cadence range 16000 - 17500 as evident in both the mean light curve (blue) and filtered cadences (red). However, overall for this period range, the density of detections is not attributable to a single systematic source.}
\label{fig:skyline_plot}
\end{figure}

\paragraph{Robust Statistic}\label{sec:robust}
A chi-squared statistic, formed from constituent elements of the total detection statistic \citep{Seader_2013}, is used in the TESS SPOC pipeline to filter false alarms. We construct a similar, although simplified, robust metric in this work, based on the principle that a true astrophysical transit should have consistent depth across all individual transit events.

We derive this robust metric for our detector following the approach of \citet{Seader_2013} but without their use of adaptive noise estimation or a multi-scale wavelet decomposition. Each transit signal $\mathbf{t}=\sum_s \mathbf{t}_s$ can be decomposed as the sum of single transit events $\mathbf{t}_s$ which are zero outside each individual transit domain. 

The detection statistic is invariant to the choice of template transit depth (Appendix \ref{sec: decompose}). For a chosen transit template $\mathbf{t}$ of fixed depth and other transit parameters, we assume the light curve contains a true transit $\mathbf{\Tilde{t}}$ described by a scaling $\alpha = \frac{\|\mathbf{\Tilde{t} \|}}{\| \mathbf{t} \|}$ of the template transit $\mathbf{t}$ of the form: $\mathbf{\hat{y}}_{i}= \alpha \sum_{s}\mathbf{t}_s + \mathbf{z}_{i}$.

From Equation~\ref{eq: detector} the single-transit detection statistic therefore has the following distribution for each hypothesis \citep{Seader_2013}:
\begin{align}
    T(\mathbf{\hat{y}}_{i}; \mathbf{t}_s|_0 ) \sim \mathcal{N}(0,1)\\
    T(\mathbf{\hat{y}}_{i}; \mathbf{t}_s)|_1) \sim \mathcal{N}(\alpha \cdot m_s, 1)
\end{align}

where $m_s=\sqrt{\mathbf{t}_s^T\Cov_z^{-1}\mathbf{t}_s}$. Given the form of these distributions and allowing $\alpha \geq 0$ to describe both the null and alternate hypothesis, the negative log likelihood of the observed single-transit detection statistics for a candidate transit $\mathbf{t}$ can be expressed in the relation:

\begin{align} \label{eq: rob}
    -\ln \left[ \mathcal{L} \left( \prod_{s} p\left(T_i ( \mathbf{\hat{y}}_{i}; \mathbf{t}_s) = \bar{\alpha} \cdot m_s \right) \right) \right] \propto \frac{1}{N_s} \sum_s^{N_s} \left( T_i ( \mathbf{\hat{y}}_{i}; \mathbf{t}_s) - \bar{\alpha} \cdot m_s \right)^2 
\end{align}

where $\bar{\alpha}$ is robustly estimated as the sample median of $\left\{ \frac{T_i(\mathbf{\hat{y}}_{i}; \mathbf{t}_s)}{m_s}\right\}$ and $p$ denotes probability. 

The offset statistic $ T_i ( \mathbf{\hat{y}}_i) - \bar{\alpha} \cdot m_s$ has a standard normal distribution, therefore the scaled sum of the variables in Equation \ref{eq: rob} is theoretically $\chi^2(N_s)$ with an expected value of $1$ \citep{Seader_2013}. A lower value of the negative likelihood indicates that the light curve is well-described by the hypothesis model for a transit (Equation~\ref{eq:hypothesis}). However the metric may exceed the theoretical distribution if the noise is non-Gaussian or if the search space discretization causes variable mismatch between the candidate and true transits. In particular shorter-period transits will be subject to greater mismatch error.

We applied this statistic to long-period signals with fewer than five transits because some of these detected transits were found to contain a small number of cadences of significant outlier noise. In this case we rejected transit signals for which the sample variance of the detection statistic across all individual transits in the signal exceeded 1.5. Furthermore over all candidate detections we rejected transits where $\hat{\alpha} < 0.0001$. In aggregate, these metrics removed $\sim 10\%$ of targets with detections remaining after filtering by the correlation statistic.

\paragraph{Eclipsing Binary Stars}

Astrophysical false alarms can also be produced by eclipsing binaries (EB) \citep{eb_fa}. A catalog of candidate eclipsing binaries\footnote{tessebs.villanova.edu} was produced from the first two years of TESS data by \citet{Prsa_2022}. We rejected candidate exoplanet detections if the target has been designated as a candidate eclipsing binary, excluding 107 of our targets. As in the search by \citet{Justesen_2021, Prsa_2022}, we produced even-odd phase-folded light curves to identify eclipsing binaries from the ratio of odd and even transit depths. The flagged eclipsing binary candidates have a ratio of odd to even transit depths in excess of 10$\%$. We use this as a coarse filtering metric here, and defer to future work the incorporation of transit depth uncertainty as used in \citet{Twicken_2018}.

\subsection{Injection Tests} \label{sec: sim}

As in \citet{taaki2020bayesian}, injection tests were performed using the joint detector and a reference standard sequential detector to assess first-order algorithm performance. Injection tests are commonly used in this context \citep{gilliland, 2013ApJS..207...35C, 2015ApJ...810...95C, 2016ApJ...828...99C, weldrake2005absence,burke2006survey,foreman}.

For these tests a total $N=1000$ light curves were randomly selected from the target selection of year 1 data from the CVZ; these comprise a significant fraction of the available data. A transit signal for a single exoplanet was injected into each selected light curve. The transit signals were simulated using the {\it transit} software library\footnote{http://dfm.io/transit} which includes limb darkening \citep{mandel2002analytic, kip1} and Keplerian orbital dynamics. The transit parameters of the injected signals were drawn from a representative exoplanet population (Table~\ref{tab:transitparams}) informed by prior population characterization \citep{foreman,kip1,kip2}. 

The reference detector was implemented as described in \citet{taaki2020bayesian}. This detector has standard detrending and detection phases. The detrending assumes no transit signal present $\mathbf{y}_i = \mathbf{s}_i + \mathbf{V}\mathbf{c}_i$ (hypothesis $H_0$ (Equation~\ref{eq:hypothesis})) and $\mathbf{V}$ is spanned by basis vectors $\mathbf{v}_k$ (Section~\ref{sec: det model}) which are here the TESS single-scale CBV (as in Section~\ref{sec:sys_noise}). The estimated basis vector coefficients $\mathbf{\hat{c}^{MAP/MMSE}_{0,i}}$ were obtained using a maximum-a-posterior (MAP) estimator (here equivalently a minimum mean square error (MMSE) estimator) and using systematic and statistical priors obtained from the injected light curves using the same procedure described in Sections~\ref{sec:stoch_noise} and \ref{sec:sys_noise}.

The estimator can be derived either by the expectation with respect to the posterior distribution $\mathbb{E}_{H_0}(\mathbf{c_i}|\mathbf{y})$, or as shown by \citet{kep},  maximizing the log of the posterior $\argmax_{\mathbf{c}_i} \ln p_{H_0}(\mathbf{c}_i | \mathbf{y}_i)$:
\begin{align}
\mathbf{\hat{c}^{MAP/MMSE}_{0,i}} = (\mathbf{V^T}\Cov_{s,i}^{-1}\mathbf{V} + \Cov_{\mathbf{c}}^{-1})^{-1} (\mathbf{V^T}\Cov_{s,i}^{-1}\mathbf{y}_i+  \Cov_{\mathbf{c}}^{-1}\mu_{\mathbf{c}_i})
\label{eq:detb_c0map}
\end{align}
The light curves were then detrended as $\mathbf{y}^{co}_{i} = \mathbf{y}_i - \mathbf{V} \mathbf{\hat{c}_{i,0}^{MAP/MMSE}}$. The sequential detection step was performed using the detection statistic (Equation~\ref{eq: detector}) decomposed over sector (Appendix~\ref{sec: decompose}) with $\Cov_{\mathbf{z},i}$ replaced by $\Cov_{\mathbf{s},i}$ as in \citet{taaki2020bayesian}. 

\begin{deluxetable}{cc}
\tablecaption{Injected Signal Parameter Distribution} 
\tablehead{
\colhead{Transit Parameter} & \colhead{Distribution} 
}
\startdata
Period $P$ (days)  & $U$(1, 100.)  \\
Radius ratio of planet to host star (\%) & $U$(0.01, 0.2) \\
Transit epoch $e$ (days) & $U$(0, $P$) \\
Impact parameter (stellar radii) & $U$(0, 1) \\
Argument of periapse $\omega$ (rad) & $U$(-$\pi$, $\pi$)\\
Limb darkening parameters: $q_1$, $q_2$ & $U$(0, 1)
\enddata
\tablecomments{The distribution of injected signal parameters. A uniform probability density function over the domain $\{x_1,x_2\}$ is denoted as $U(x_1,x_2)$. See \citet{kip1} for a definition of limb-darkening parameters. }
\label{tab:transitparams}
\end{deluxetable}

No post-processing or filtering was performed for the injection tests because their purpose was solely to characterize the comparative performance of the joint and sequential detection approaches. The detection statistics are  implemented as in Section \ref{sec: transit_search}, allowing candidate transits over a sector to shift by $\pm$ 2 h. A candidate detection was recorded for an injected light curve if the detection statistic exceeded the threshold $T(\mathbf{\hat{y}}_i) > 10$ for a candidate transit signal $\mathbf{t}$ and considering only the detected candidate transit signal $\mathbf{t}$ with the maximum test statistic. The number of light curves with detections is denoted $N_d$. The true injected transit signal $\mathbf{t}_{inj}$ in a light curve was considered correctly recovered the candidate transit signal $\mathbf{t}$ if either: i) the orbital periods of the detected $P(\mathbf{t})$ and injected $P(\mathbf{t}_{inj})$ transit signals matched within $|P(\mathbf{t})-P(\mathbf{t}_{inj})| < 2$ h; or, ii) the correlation metric between the injected and detected transit signals exceeded 0.75. As described in Section \ref{sec: transit_search}, the transit template is allowed to shift $\pm 2$ hr and this was similarly applied when computing the correlation metric. Furthermore a box approximation was used for both injected and detected transits. Criterion (ii) was added because for long-period ($>$ 15 d) injected transit signals a significant proportion of the transits may occur on data gaps. A detected transit may then recover the majority of transit events however may not match in detected period per criterion (i). Since the quasi-false alarm rate is intended to provide a measure of erroneous detections from systematics we do not count detections which are correlated with the injected transit as false alarms. A correlation threshold of 0.75 was chosen as this implies that more than half of detected and injected individual transits match.

 The number of injected light curves with correct detections meeting these criteria is denoted $N_c$. The detection rate is $R_D=\frac{N_c}{N}$. The quasi-false-alarm rate $R_{QFA}=\frac{N_d-N_c}{N}$ measures the rate of incorrect detections. We note that a false alarm may in fact be a true astrophysical transit present in the TESS light curve before the injected transit signal was added; this particularly affects detection rate estimates for weaker injection transits. However, this is mitigated by the low expected density of preexisting transits in the set of injected light curves and the higher strength of injected transits in general relative to true preexisting transits. Therefore we believe that the quasi-false-alarm rate $R_{QFA}$ reasonably identifies noise properties or features that pose challenges to correct detection.

\section{Results}\label{sec:results}

\subsection{Injection Test Performance} \label{sec: inj_result}
The injection tests used to assess algorithm performance are described in Section~\ref{sec: sim}. In these tests the standard sequential detector used as a reference achieved a detection rate $R_D = 79.8\%$ and a quasi-false-alarm-rate $R_{QFA}=19.1\%$. The joint detector \citep{taaki2020bayesian} as implemented here, achieved a marginal increase in detection rate of $0.2\%$, although this is not statistically significant given the sample size. The detection rate $R_D$ is plotted in Figure~\ref{fig: roc} against quasi-false-alarm rate $R_{QFA}$ as a function of detection threshold $\tau$ for both the joint and the standard reference algorithms. The detection rate of the joint detector at $\tau=10$ is shown in Figure~\ref{fig:det_eff_s_y1}, tabulated over orbital period and the planet-to-star radial ratio $\frac{R_p}{R_*}$ of the injected signals.

For the injection tests, the distribution of quasi-false alarms and correct detections over detected period is shown in Figure~\ref{fig:fa_distr}. Over 60 $\%$ of the quasi-false alarms occur due to unmodeled systematics or scatter spanning the $13.7\ \rm{d}$ downlink cycle gap and also possibly associated with the 2-6 day momentum dump cycle. In addition, the quasi-false-alarm rate increases with period; this is likely due to outlier events remaining in the data.  

\begin{figure}[ht!]
\centering
\includegraphics[width=.5\linewidth]{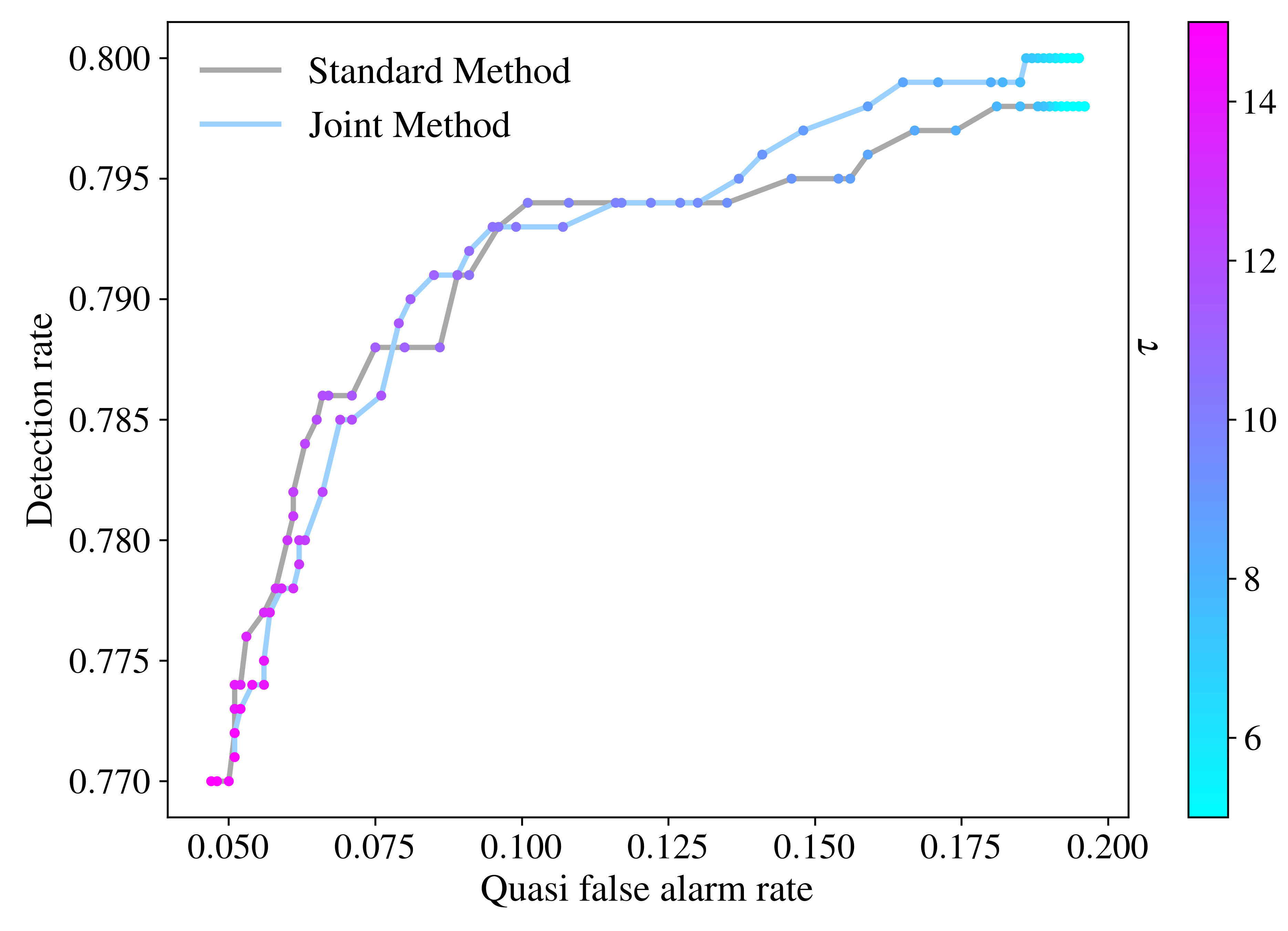}
\caption{Injection test detection rate versus quasi-false alarm rate over detection threshold $\tau$ (right color wedge) for the joint and standard reference detection algorithms. The two methods are comparable in performance, showing only a marginal improvement in detection power for the joint method.}
\label{fig: roc}
\end{figure}

\begin{figure}[ht!]
\centering
\includegraphics[width=.45\linewidth]{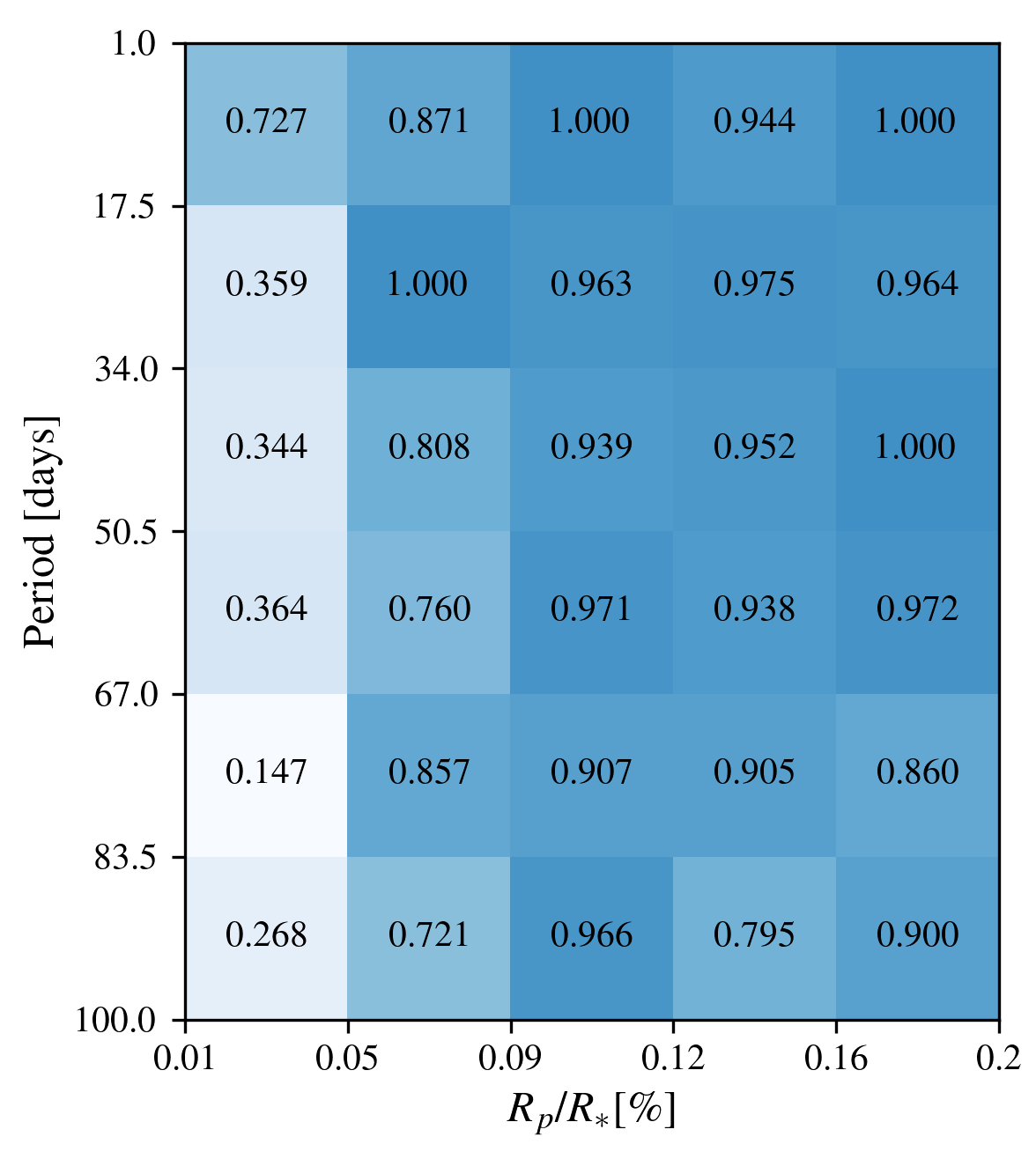}
\caption{The detection rate of the joint detector at a threshold of $\tau = 10$ for transit injection tests with 1000 light curves from year 1, tabulated over orbital period and planet-to-star ratio $\frac{R_p}{R_*}$. The distribution of simulated transit signal parameters can be found in Table~\ref{tab:transitparams}. The mean detection rate and quasi-false-alarm rate are $R_D = 80.0\%$ and $R_{QFA}=19.1\%$ respectively.}
\label{fig:det_eff_s_y1}
\end{figure}

\begin{figure}[ht!]
\centering
\includegraphics[width=.5\linewidth]{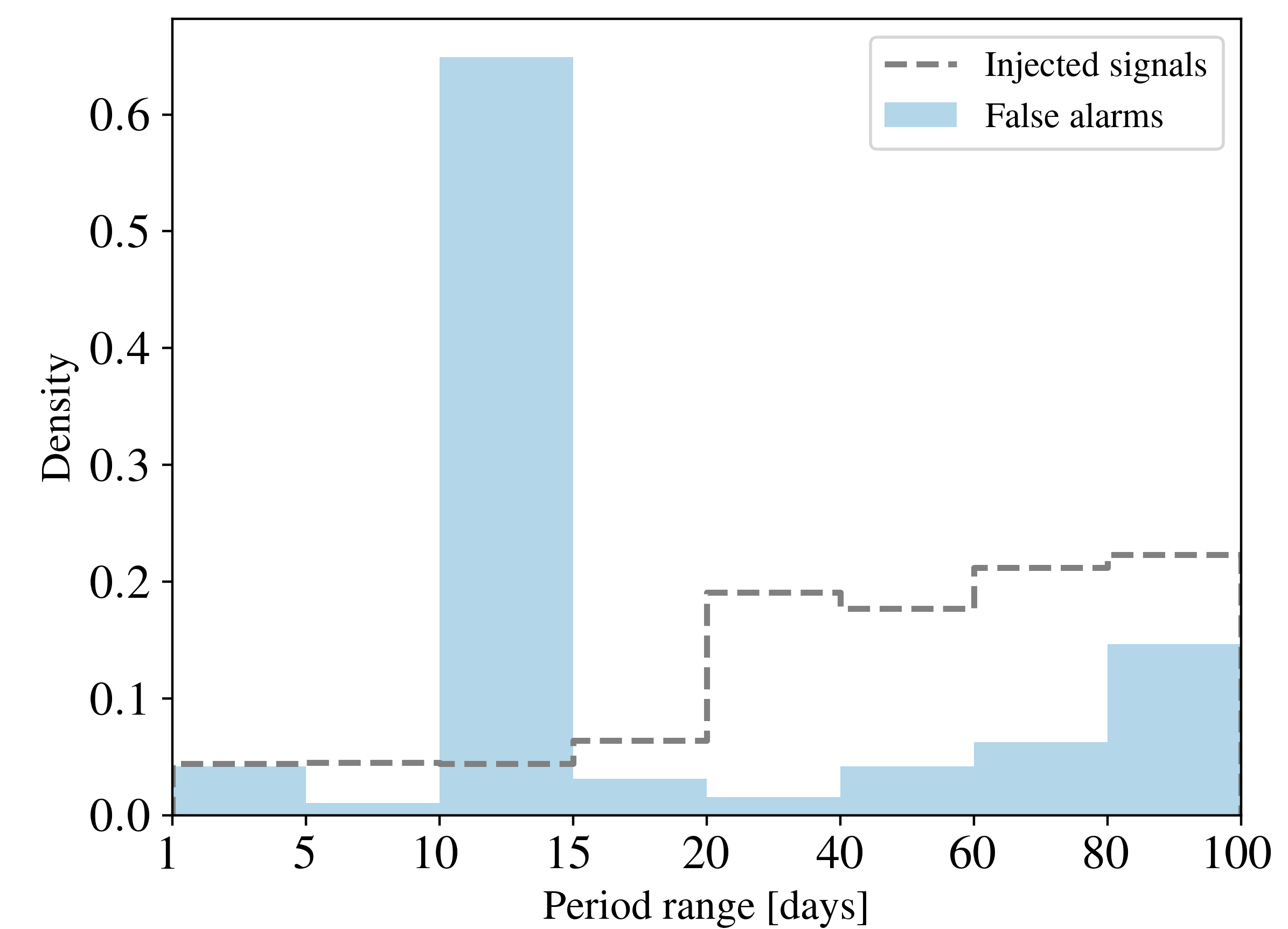}
\caption{For the injection tests, the distribution of quasi-false-alarm detections (188/1000) (blue) and injected transits (gray dashed) as a function of their detected or injected periods respectively
plotted as a normalized density. Half of false-alarm detections have a period between 10-15 d, likely due to the 13.7 d downlink cycle. The density of false alarms further increases with period, likely due to outlier scatter. We note that the bin size on the x-axis is not uniform.}
\label{fig:fa_distr}
\end{figure}

\subsection{TESS Search} \label{sec: tess_search}
An important goal of this work is to perform an independent search of the first three years of the TESS 2-min SAP light curves from the CVZ using the joint detector \citep{taaki2020bayesian} as implemented here. No light curves with known TESS Objects of Interest (TOI), produced by the SPOC/QLP pipelines, were excluded from our search in advance. We compare our recovery rate of known TOI as a coarse form of validation. We note that the period range searched by the SPOC/QLP pipeline differs from ours, and by definition, any candidate transits lying outside of our search range are excluded from the listed TOI here. 
Post-processing and filtering (Section~\ref{sec: postprocess}) were applied to our candidate detections, which were obtained at a detection threshold $\tau = 10$. The detection threshold was set based on the injection tests balancing the false alarm rate against sensitivity to weak transit signals. As noted in Section~\ref{sec: postprocess}, post-processing includes the removal of known or suspected eclipsing binaries (EB). A total of 78 candidate detections were removed as known EB cataloged in \citet{Prsa_2022}, including three identified by SPOC and one listed in \citet{triple_eb}. We note that a further 21 known EB are cataloged for light curves in which we did not detect any transits. The majority of missed EB have very short orbital periods $< 2 $ d, with a large morphology parameter $c > 0.5$ \citep{Matijevi_2012}. These high-contact systems have phase-folded light curves with wide eclipses, unlike transiting exoplanets. 

After post-processing and filtering, approximately $28 \%$ of light curves contained a detection, with $551/1851$ in year 1, $659/2046$ in year 2 and $420/1922$ in year 3. The distribution of these detections over period is shown in Figure~\ref{fig:det_short_overall} for each year separately. This Figure confirms the quasi-false alarm distribution with period found in the injection tests (Figure~\ref{fig:fa_distr}). The post-processing and filtering is somewhat robust to outlier signals, however these processes are not robust to unmodeled quasi-periodic systematics that may mimic transit signal. This is particularly true for those associated with the 13.7 d downlink and 2-6 d momentum dump cycles, as suggested in these Figures. To explore this further, the detection count in the TESS data search is tabulated in Figure~\ref{fig:det_pd_short} over detected period $P$ (for $P < 19\ {\rm d}$) and detection statistic value $10 \leq T(\mathbf{\hat{y}}) \leq 20$ for all three years of TESS data combined. This provides further support for this hypothesis. Further supporting evidence is provided by Figure~\ref{fig:downlink_scatter} which shows a strong correlation between detected transits and sector downlink edges and also shows that these are regions of higher systematic light curve scatter.

Approximately $63 \%$ of known TOI were recovered in these detections as described below in Section \ref{sec: TOI}. Informed by the quasi-false alarm distribution over period discussed above, the majority of exoplanet candidates with periods between $10-15$ days were excised during this manual vetting. Similarly there was careful vetting of periods in the 2-6 d interval due to the momentum dump cycle and the overlapping 1-5 d interval quasi-periodic stellar variability known in the 1-5 d interval from Kepler TCE processing \citep{Thompson_2018}. Additional manual vetting of phase-folded light curves was performed. However, MAST SPOC Data Validation reports indicated that the remaining candidate detections were most likely eclipsing binaries not previously identified as threshold crossing events.

\begin{figure}[ht!]
\centering
\includegraphics[width=.5\linewidth]{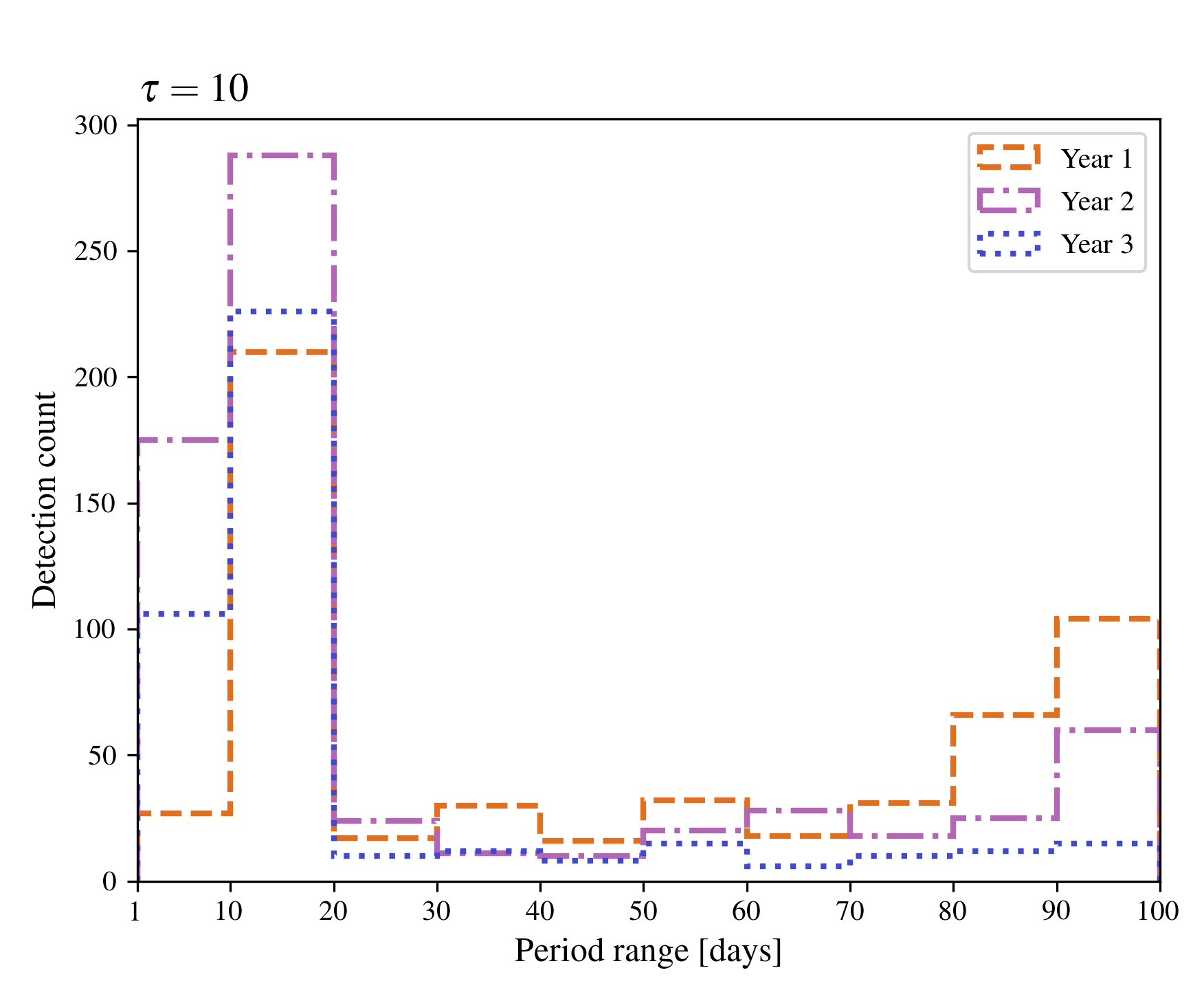}
\caption{The distribution of detections in the TESS data search (Section~\ref{sec: tess_search})  each year over period. As in the injections tests (Figure~\ref{fig:fa_distr}) there is an increase in suspected false alarm detections for periods $P < 20\ \rm{d}$ likely due to residual systematic scatter.}
\label{fig:det_short_overall}
\end{figure}

\begin{figure}[ht!]
\centering
\includegraphics[width=.4\linewidth]{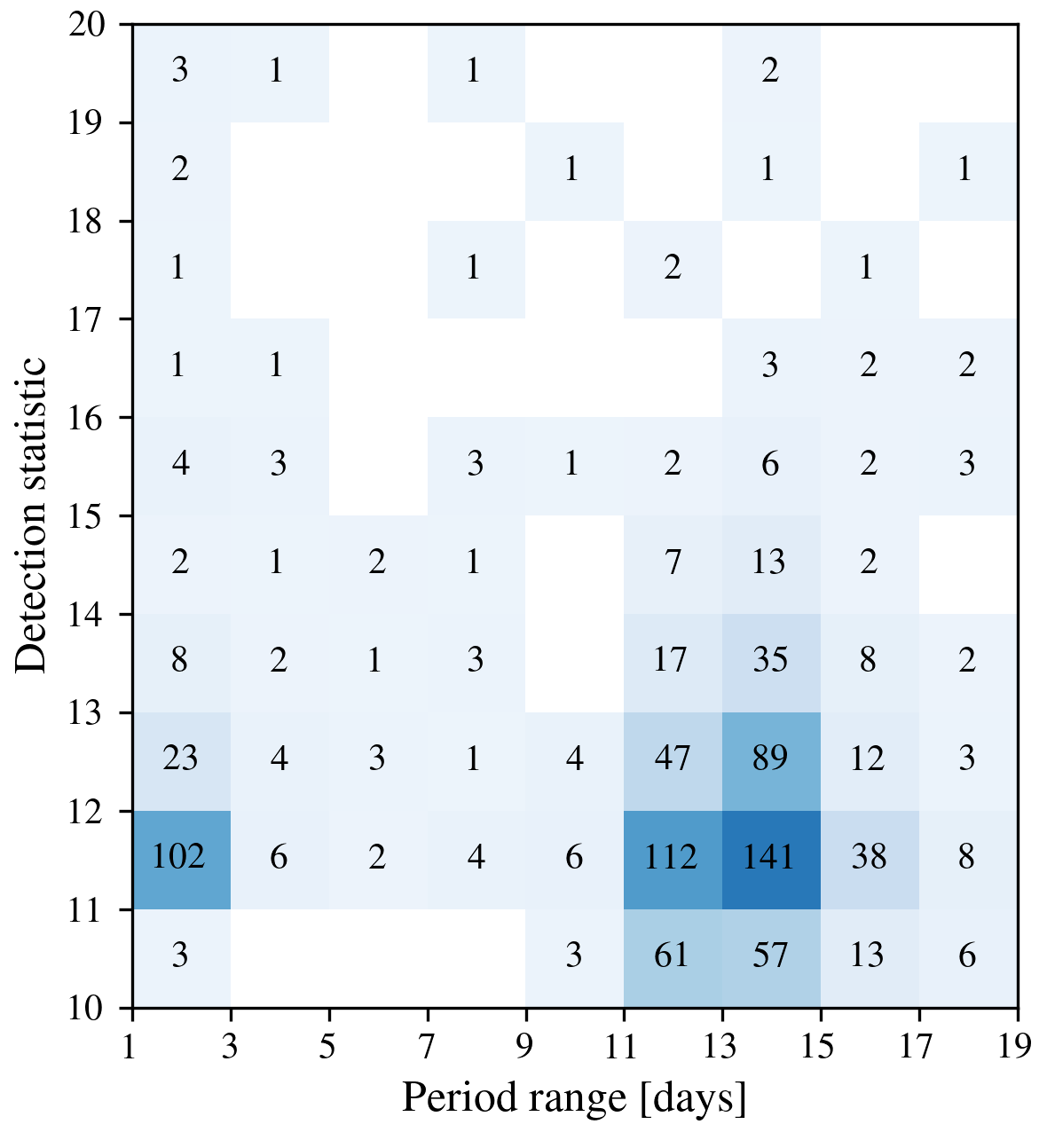}
\caption{Detection counts for the TESS data search tabulated over detected period ($P \leq 19\ {\rm d}$ and detection test statistic $10 \leq T(\mathbf{\hat{y}}) \leq 20)$) for all three years combined. The overdensity of detections between 1-3 days and 11-15 days are very likely false alarms due to residual systematic scatter. Both period ranges map directly to mission instrumental cycles, as described in Section \ref{sec: inj_result}.}
\label{fig:det_pd_short}
\end{figure}

\begin{figure}[ht!]
\centering
\includegraphics[width=.5\linewidth]{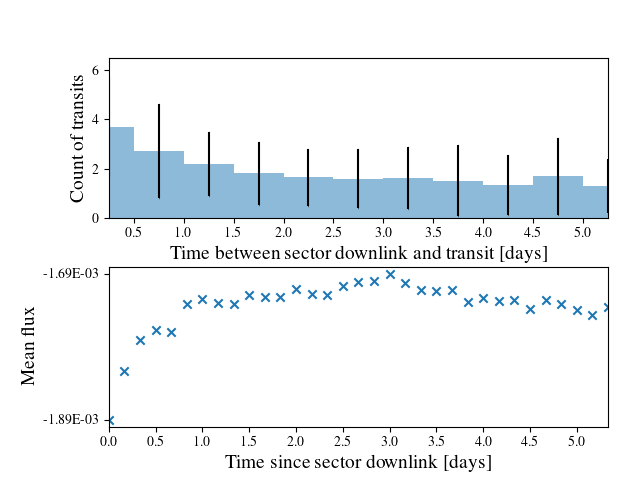}
\caption{In the upper panel, the count of transit events binned in 0.5d time-offset intervals between the sector downlink event and each transit event, for year 2 of detected transits and including all sectors. The error bar represents the standard deviation of the transit count over the set of light curves. In the lower panel, the median light-curve flux density in the same time-offset bins is shown at bin width 4 h. Increased negative scatter occurs close to the sector downlink times and is correlated with an increased transit event count, suggesting the many detections with periods between 10-15 d are likely false alarms.}
\label{fig:downlink_scatter}
\end{figure}

\subsubsection{TOI Recovery}\label{sec: TOI}

TESS project TOI are produced by filtering threshold crossing events (TCE) identified by the QLP/SPOC pipeline. Automated and manual vetting is applied in this process to reduce false alarms and false positives \citep{toi_guerrero}. We downloaded the project TOI list from the ExoFOP database\footnote{\url{https://exofop.ipac.caltech.edu/tess/}}. As noted under the TOI release notes\footnote{\url{https://tess.mit.edu/toi-releases/}} there may be inconsistencies between the fitted transit parameters across the TOI lists published by the TESS Science Office (TSO), MAST, or ExoFOP. We note that only a small fraction $(\sim 12 $\%$)$ of TOI  are, as yet, confirmed exoplanets \citep{melton2023diamante}, as reasonably expected. 

Given this heterogeneity, we used weaker criteria for TOI recovery than those used in the injection tests (Section~\ref{sec: sim}). For each light curve we consider a detected transit signal $\mathbf{t}$ to recover a known TOI $\mathbf{t}_{TOI}$ associated with the target if: i) the detection statistic for $\mathbf{t}$ exceeds the detection threshold $\tau = 10$; ii) the detection statistic for $\mathbf{t}$ is the maximum for the light curve; and, iii) the period of the detection $P(\mathbf{t})$ matches the period of the TOI $P(\mathbf{t}_{TOI})$ within a tolerance $|mP(\mathbf{t})-P(\mathbf{t}_{TOI})| < 2\rm{h}\ (m \in \{2,3,\frac{1}{2},\frac{1}{3}\})$; the integer multiplicity in allowed period match addresses a small number of cases $(<5)$ in which the detected or TOI periods may have a low-order integer ambiguity due to data gaps, as noted for the QLP \citep{toi_guerrero}. These three criteria define the TOI {\it match} or {\it recovery} rate in what follows. If the detection statistic for $\mathbf{t}$ is above the threshold but not a maximum (i.e. only (i) and (iii) are satisfied) then this defines the TOI {\it recall} rate as used here. We note that 12 light curves contained more than one TOI, we count these as a single TOI in the total TOI count, a singular recall if any TOI are among the detections, and a single match if any of the TOI are recovered according to the detection criteria. 

Over the three years of data we found a $63\%$ TOI recovery and a $73\%$ TOI recall rate. Table~\ref{tab:toi} shows for each year the number of light curves with known TOI, the recall and match rates for these TOI, and a separate match rate excluding TOI designated by TESS as potential false positives (FP) or false alarms (FA). The TOI count and matched TOI detections is shown over period in Figure~\ref{fig:toi}. This Figure shows a slight improvement in TOI recovery for periods between 10-100 d.

\begin{figure}[ht!]
\centering
\includegraphics[width=.5\linewidth]{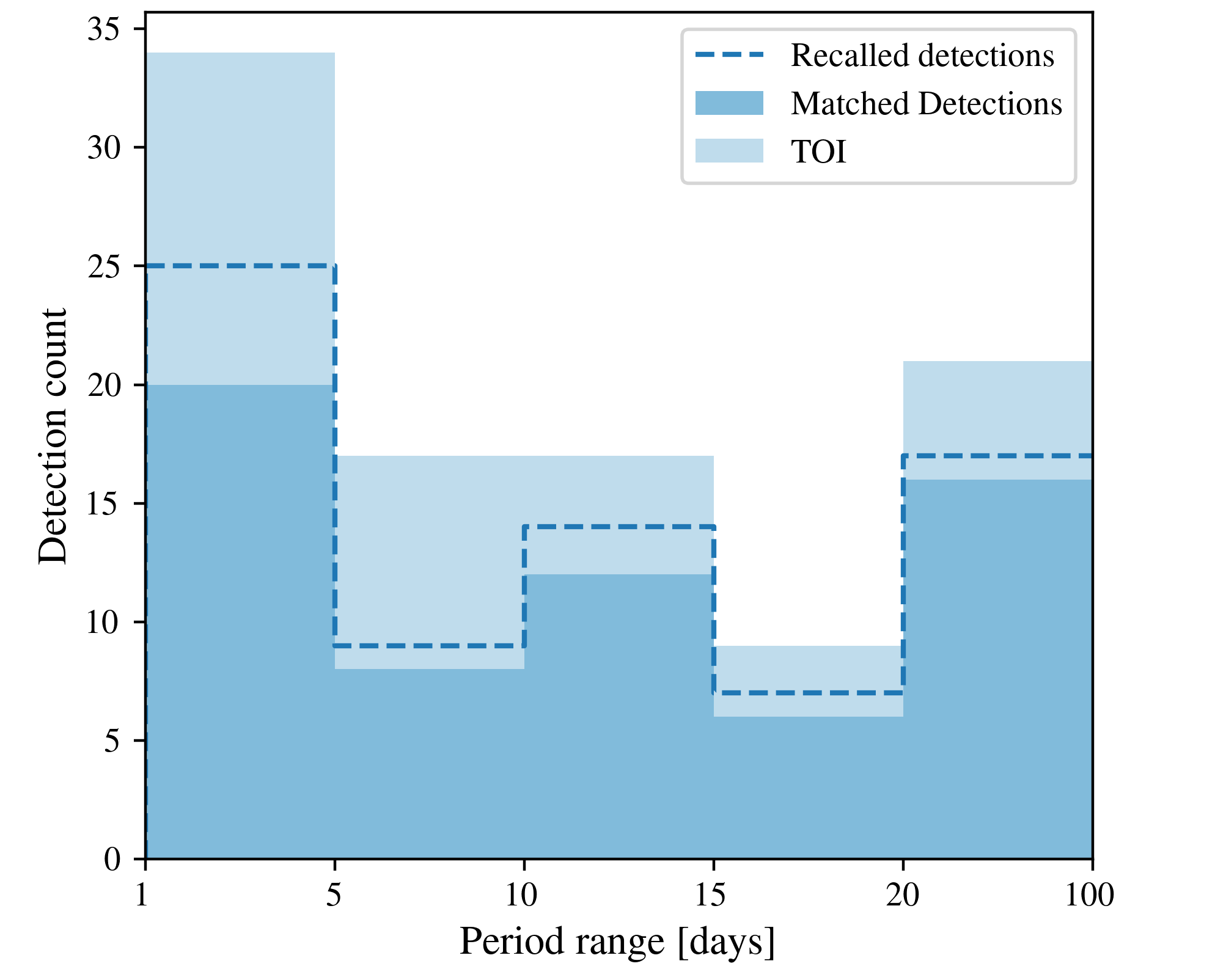}
\caption{The distribution of known TOI (light blue), known TOI recovered as the top detection by the joint detector (darker blue) and known TOI recovered as any detection above threshold by the joint detected (dashed line) over period. The TOI recovery rate increases slightly in the period range 10-100 d. We note that the bin size on the x-axis is not uniform.}
\label{fig:toi}
\end{figure}

\begin{deluxetable}{ccccccc}[ht!]
\tablecaption{TOI detections at $\tau = 10$ \label{tab:toi}}
\tablehead{
 \colhead{} & \colhead{Total TOI} & \colhead{TOI recalls}  & \colhead{Recall rate} & \colhead{TOI matches } & \colhead{Match rate} &\colhead{Match rate ($\overline{{\rm FP}},\overline{{\rm FA}}$)}}
\startdata
Year 1  & 30 & 20 & 67 \% & 17 & 57 \%  & 55 \% \\ 
Year 2 & 21 & 18 & 86 \% & 15 & 71 \%   & 68 \% \\
Year 3  & 47 & 34 & 72 \% & 30 & 64 \%  & 64 \% \\
Overall & 98 & 69 & 73 \% & 62 & 63 \%  & 62 \% \\
\enddata
\tablecomments{From left to right the columns are: year number, the total number of light curves with TOI present, the number of TOI recalled, the associated recall rate, the number of TOI matches or recoveries, the associated match rate, and the match rate with any TOI that are designated by the TESS project as false positives (FP) or false alarms (FA) excluded. The final row are the combined statistics across years 1 to 3.}
\end{deluxetable}

\section{Discussion}\label{sec:discussion}
A goal of the current work is to evaluate the feasibility of applying our joint detector, previously developed and evaluated with Kepler data \citep{taaki2020bayesian}, to data from the TESS mission: another wide-field exoplanet transit survey with different instrumental systematics and observing cadence. We find the modifications and enhancements required to be algorithmically feasible (Section~\ref{sec: implementation}) and computationally tractable (Appendix~\ref{sec: compute}), albeit remaining in the HPC domain \citep{taaki2020bayesian}.

An additional goal of the paper is to assess the statistical performance of the joint detector in the new application to TESS data. In transit injection tests (Section~\ref{sec: inj_result}) we find the joint detector to be comparable to or slightly exceed (Figure~\ref{fig: roc}) the statistical performance of a standard sequential detrending and detection algorithm, used here as a reference.  At a detection threshold $\tau=10$, the joint detector achieved a detection rate $R_D=80.0\%$ at a quasi-false-alarm rate $R_{QFA}=19.1\%$ for $N=1000$ injected transit signals. For fixed $R_{QFA}$ this is a $\triangle R_D=0.2\%$ improvement over the reference sequential detector; however this is not statistically significant. We note an important caveat that the standard sequential detector (Section~\ref{sec: sim}) is implemented as a modification of our pipeline and is only representative of this class of sequential detector; we make no claim of optimality relative to other implementations. 

In prior work with Kepler data \citep{taaki2020bayesian} the joint detector was found to have an advantage in detecting short-period, low transit-depth exoplanets relative to a standard sequential detector, likely by mitigating overfitting. In the current work with TESS data, no such advantage is evident. In this work we have used the well-constrained TESS project CBV for all detectors, thereby possibly minimizing overfitting by the sequential detector. The standard deviation of each cadence under the systematics noise prior is of the comparable magnitude to the stochastic noise, which includes stellar variability (Figure~\ref{fig:std_flux}). In addition, the single most likely sources of residual systematic error and therefore false alarm are the 13.7 d instrument downlink cycle and the 2-6 d momentum dump cycle, as discussed in further detail below. This effect is sufficiently prominent it may have overshadowed more subtle detection differences evident in the prior Kepler work \citep{taaki2020bayesian}, limiting our conclusions about joint detection here. Our injection tests provide insight into the algorithmic choices for these specific data. They do not necessarily generalize to other datasets or address the global optimality of joint modeling. This question is broad \citep{foreman, kov_joint, Garcia_2024} but remains unsettled in the literature. Notably, \citet{Garcia_2024} identify a benefit of their approach compared to other common transit search implementations for rapidly rotating M dwarfs. 

We find the statistical error of the joint detector applied to TESS data to be strongly dependent on exoplanet orbital period $P$. Specifically, the transit injection tests show a significantly enhanced quasi-false-alarm rate for the detected orbital period range $P \in [10,15]\ \rm{d}$ and a secondary enhancement at $P < 5\ \rm{d}$ (Figure~\ref{fig:fa_distr}). The full search of the input TESS data (Section~\ref{sec: tess_search}) similarly show (Figures~\ref{fig:det_short_overall} \&~\ref{fig:det_pd_short}) an anomalous excess of detections in the same period range, with a secondary detection excess in the range $P < 5\ \rm{d}$, both likely quasi-false alarms (Figure~\ref{fig:fa_distr}). There is evidence that unmodeled systematics near the 13.7 d instrument downlink data gap are responsible for the excess detections for $P \in [10,15]\ \rm{d}$ \citep{Kunimoto_2023}, and also as examined in the analysis in Figure~\ref{fig:downlink_scatter}. Furthermore, the spurious detections for $P \in [2,5]\ \rm{d}$ likely occur due to unmodeled systematics associated with the momentum dump cycle, also noted as a source of false positives in the TESS TOI search \citep{toi_guerrero}. The robust CBVs used do not include edge flux ramps prior to downlink, as these are typically corrected via spline regression in an independent part of the SPOC pipeline \citep{kep_handbook}. Edge correction may help to mitigate the density of these spurious detections. However, the SPOC pipeline also produced increased false alarms between $P \in [10,15]\ \rm{d}$ \citep{fausnaugh2018tess} due to downlink scatter. The sensitivity of statistical error to exoplanet period is not specific to the joint detector. It occurs with other transit detectors, such as the reference sequential detector used in this work, and is foundationally related to the TESS observing and operations schedule structure.

The injection tests (Figure~\ref{fig:fa_distr}) also show an increase in the quasi-false-alarm rate for longer periods $P > 60\ \rm{d}$, as also shown in the full TESS data search (Figure~\ref{fig:det_short_overall}). We believe that these false-alarm detections occur due to non-periodic instances of outlier scatter, particularly for the year 1 data (on which the transit injection tests were performed). This likely reflects an inadequacy in our stochastic noise estimator, specifically a failure to model local noise variability (Section~\ref{sec: postprocess}) and the assumption of stationarity over a sector. This suggests that future work to incorporate local downlink and outlier scatter in a stochastic noise prior using window estimation akin to \citet{jenkins_2010} may be a productive approach in future work. Furthermore, when combining multi-sector transit detection statistics, the $\pm 2$h epoch shift tolerance may contribute instances of false alarms with longer orbital periods. 

The input light curves in the full TESS search (Section~\ref{sec: tess_search}) contain 98 unique targets with known TOI (Table~\ref{tab:toi}), provided by various detection pipelines including SPOC \citep{spoc} and the QLP \citep{toi_guerrero}. Under the definitions and criteria in Section~\ref{sec: TOI}, recall rate of $73\%$ and a recovery rate of $63\%$ are broadly comparable to the TOI recovery rate of other detectors for which this metric has been evaluated \citep{chakraborty2020hundreds, eisner}. A priori, lossless TOI recovery is not expected given differences in algorithm approach, implementation, and the sources of false alarm identified by the diagnostic tests described in Section~\ref{sec: tess_search}. In particular, although the detection thresholds are not directly comparable quantities, our detection threshold $\tau = 10$ is higher than the project multiple-event-statistic (MES) detection threshold of $7.1$ \citep{Fausnaugh_2021, toi_guerrero}.
The distribution of TOI recovery rate over period $P$ (Figure~\ref{fig:toi}) shows higher TOI recovery and recall rates for period $P > 10\ \rm{d}$ and also greater scatter in these rates for $P < 10\ \rm{d}$. In contrast, the injection tests show a monotonic decrease in recovery rate with increasing period (Figure~\ref{fig:det_eff_s_y1}). This discrepancy is likely explained by the sample signal statistics of the TOI. As shown in Figure \ref{fig:missed_toi}, the TOI with orbital period $P < 10\ \rm{d}$ missed by the joint detector generally have a lower median planet signal-to-noise reported on ExoFOP compared to both the detected TOI and the missed TOI with longer orbital periods. The missed TOI have an overall median planetary SNR of 11.3, compared to the detected TOI of 23.0. 

\begin{figure}[ht!]
\centering
\includegraphics[width=.5\linewidth]{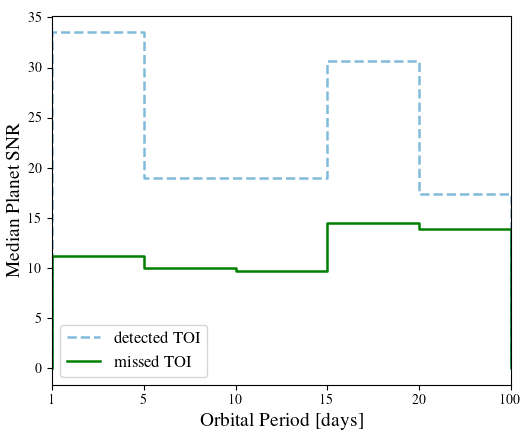}
\caption{The distribution of the median planet signal-to-noise (SNR) of missed TOI (green line) and detected TOI (dashed blue line) over period, as obtained using the joint detector. The planet SNR are those provided by ExoFOP. Although the detection thresholds cannot be compared directly, our detection threshold ($\tau = 10$) exceeds the MES detection threshold of $7.1$, and this may reasonably contribute to differences in candidate detections. In general the missed TOI have a lower planetary SNR of 11.3, compared to a value of 23.0 for the detected TOI.}
\label{fig:missed_toi}
\end{figure}

A final goal of the current work is to conduct an independent search of the TESS 2 min SAP light curves for the CVZ using the joint detector. Complementary exoplanet transit searches of these data are important to find possible missed exoplanet detections. After manual vetting to eliminate eclipsing binaries, stellar harmonics, or noise features mimicking transits and cross referencing against the MAST SPOC DV reports for threshold crossing events (TCE) we find none of the detected signals can be unambiguously classified as exoplanets. We note in closing that the results presented here depend on specific tunable parameter or implementation choices such as transit template search space discretization and chosen noise priors, amongst others. This work was not designed to fully optimize these choices. However, as noted in the Discussion, existing work including the injection tests has identified several aspects of the pipeline that could be optimized to improve performance. As noted above, a specific example includes improved systematic and stochastic noise priors to better model downlink scatter and local noise variance.

\section{Conclusions}\label{sec:conclusion}
In this work, we applied a prior joint detection method for exoplanet transits \citep{taaki2020bayesian}, which uses Bayesian priors for systematic and stochastic noise and was previously applied to Kepler data, to the first three years of TESS 2 min SAP data from the CVZ.  An exoplanet search was conducted over periods between 1-100 days. 
We modified the detector implementation to address unique aspects of TESS data and to provide an efficient HPC implementation. The detector pipeline is robust to spurious harmonics occurring due to stellar variability and gap-filling or stitching is not explicitly required. We utilized several refined post-processing methods to suppress false alarms due to outlier noise and residual quasi-periodic systematics.
\par Our resulting pipeline performance was first evaluated using transit injection tests. The joint detector was evaluated against a sequential detrending and detection algorithm used as a reference standard implementation. We also evaluated the recovery of known TOI present in the light curves by performing a full search on the TESS input data. The principal conlusions of this work are listed below.
\begin{itemize}
    \item The detector was successfully modified and applied to a new wide-field exoplanet transit mission (TESS). In optimized form it is computationally tractable but remains in the HPC domain.
    \item The injection tests show a high  detection rate $R_D=80.0 \%$ with a quasi-false-alarm rate $R_{QFA}=19.1 \% $ at a detection threshold $\tau = 10$. A marginal improvement in detection performance $\triangle R_D=0.2 \%$ is achieved relative to the reference sequential detector. We posit that the similar performance is primarily due to a systematic noise prior that is fairly well-constrained relative to stochastic variability, per light curve. 
    \item An excess of quasi-false-alarms is found in period ranges $P \in [10,15]\ \rm{d}$ and $P < 5\ \rm{d}$, most likely due to unmodeled sysmatics related to the 13.7 d downlink cycle and the 2-6 d momentum dump cycle intrinsic to the TESS operational schedule.
    \item In our full-scale search of three years of TESS 2-min light curves from the CVZ we recover $63 \%$ of known TOI with a matching orbital period and achieve a $73 \% $ recall rate for TOI when compared to all detected transits. 
\end{itemize}

\begin{acknowledgments}
\section{Acknowledgements}
We thank the anonymous referee for their insightful comments on this topic. 

This research is part of the Blue Waters sustained-petascale computing project, which is supported by the National Science Foundation (awards OCI-0725070 and ACI-1238993) and the state of Illinois. Blue Waters is a joint effort of the University of Illinois at Urbana-Champaign and its National Center for Supercomputing Applications. 

Some of the data presented in this paper were obtained from the Mikulski Archive for Space Telescopes (MAST) at the Space Telescope Science Institute. The specific observations analyzed can be accessed via \dataset[https://doi.org/10.17909/t9-nmc8-f686]{https://doi.org/10.17909/t9-nmc8-f686}. STScI is operated by the Association of Universities for Research in Astronomy, Inc., under NASA contract NAS5–26555. Support to MAST for these data is provided by the NASA Office of Space Science via grant NAG5–7584 and by other grants and contracts. 
This paper includes data collected with the TESS mission. Funding for the TESS mission is provided by the NASA Explorer Program. STScI is operated by the Association of Universities for Research in Astronomy, Inc., under NASA contract NAS 5–26555. 

This research has made use of the Exoplanet Follow-up Observation Program (ExoFOP; DOI: 10.26134/ExoFOP5) website, which is operated by the California Institute of Technology, under contract with the National Aeronautics and Space Administration under the Exoplanet Exploration Program.

\software{This work used the following external packages: Astropy \citep{astropy:2013, astropy:2018, astropy:2022}, Scikit-learn \citep{scikit-learn}, Matplotlib \citep{Hunter:2007}, NumPy \citep{harris2020array}, SciPy \citep{2020SciPy}, and transit v0.3.0 (\url{https://github.com/dfm/transit}).  }

\end{acknowledgments}

\appendix
\section{List of symbols} \label{ap: symbols}
A list of important symbols and mathematical notation used in the Appendix is provided in Table~\ref{tab:symbols}.
\begin{table}[htbp!]
\caption{List of Symbols}
\begin{center}
\begin{tabular}{r l p{10cm} }
\toprule
$\mathbf{\hat{y}}$ & Light curve \\
$\mathbf{z} \sim \mathcal{N}(0, \Cov_\mathbf{z})$ & Joint noise model \\
$i \in I$ & Index set of unique targets \\
$q \in Q$ & Set of TESS observational sectors \\
$N$ & Length of a light curve \\
$N_q$ & Length of a single sector of a light curve\\
$d \in D$ & Set of discrete transit durations \\
$P$ & Transit orbital period \\
$e$ & Transit epoch \\
$\alpha$ & Relative transit depth \\
$\delta P$ & Discrete sampling increment in period \\
$\delta e$ & Discrete sampling increment in epoch \\
$\mathbf{t} \in \mathbf{T}$ & Transit template search space \\
$\mathbf{T}_d$ & Subspace of $\mathbf{T}$ for a transit duration $d$ \\
$\mathbf{T}_{d(q)}$ & Subspace of $\mathbf{T}$ for a transit duration $d$ and a single sector of light curve data \\
$\mu_{\mathbf{t}}$ & average number of transit events per $\mathbf{t} \in \mathbf{T}$\\
\bottomrule
\end{tabular}
\end{center}
\label{tab:symbols}
\end{table}

\section{Sector-to-sector Detector} \label{sec: decompose}

\par Analogous to the discussion in \citep[Appendix B]{taaki2020bayesian}, we assume independent noise in each of the 13 sectors in each year-long light curve. Dropping the light curve index $i$, systematics are estimated sector to sector and a sector-long model for statistical noise $\mathbf{s}$ is also appropriate relative to transit timescales. For each sector $q \in Q$ the joint noise covariance is therefore denoted $\Cov_{\mathbf{z}(q)} =  \Cov_{\mathbf{s}(q)} + \mathbf{V}_{q} \Cov_{\mathbf{c}(q)} \mathbf{V}_{q}^T$.

Scaling the candidate transit signal $\mathbf{t}$ by $\alpha \in \mathbb R^+$ will not affect the matched filter statistic \citep{kay1998fundamentals}. 
For a signal $\bold{t}_{\alpha} = \alpha \bold{t}$ :
\begin{align} \label{eq: det2}
    \frac{\bold{\hat{y}}^T\mathbf{\Cov_{z}}^{-1}\bold{t}_{\alpha}}{\sqrt{\bold{t}_{\alpha}^T \mathbf{\Cov_{z}}^{-1} \bold{t}_{\alpha}}} =    \frac{\bold{\hat{y}}^T\mathbf{\Cov_{z}}^{-1}\alpha\bold{t}}{\sqrt{\alpha^2 \bold{t}^T \mathbf{\Cov_{z}}^{-1} \bold{t}}} =    \frac{\bold{\hat{y}}^T\mathbf{\Cov_{z}}^{-1}\bold{t}}{\sqrt{\bold{t}^T \mathbf{\Cov_{z}}^{-1} \bold{t}}}
\end{align} 
Therefore a candidate transit depth is not needed to parameterize $\mathbf{t}$. 
However, to correctly compute a year-long detection test each sector light curve $\mathbf{\hat{y}}_{q}$ should be normalized so that transit strength is constant across all sectors. 
In our implementation, we chose to compute detection tests per sector and approximate a year-long detection test as the normalized sum of sector-wise matched filters described in Equation \ref{eq: final_det}. This average of detection tests is equivalent to a year-long detection test with the same noise model if the sector-to-sector noise $\mathbf{\Cov}_{\mathbf{z}{(q)}}$ is identically distributed between sectors. In informal tests, this detection statistic is found to be marginally lower, as the relative contribution of sectors with lower estimated noise levels is reduced.

\begin{align} \label{eq: final_det}
  T(\mathbf{y}) = \frac{1}{\sqrt{Q}} \sum_{q \in Q} \frac{ \mathbf{\hat{y}}_{q}^T \mathbf{\Cov}_{\mathbf{z}(q)}^{-1} \bold{t}_q}{\sqrt{\bold{t}_q^T \mathbf{\Cov}_{\mathbf{z}(q)}^{-1} \bold{t}_q}} \LRT{H_1}{H_0} \tau
\end{align}

\subsection{Computational methods and complexity} \label{sec: compute}

This Section characterizes the computational complexity of the detection test kernel in the pipeline implementation (Figure \ref{fig: pipeline_overview}). 
This work was completed with a $250,000$ node hour allocation on the Blue Waters petascale system, where one node hour equals $32$ core hours. For each year of light curve data, the detection tests required $\sim$ $50,000$ node hours of compute time and used $\sim50$ TB of intermediate storage.

\subsubsection{Efficient computation of multiple transit detection tests}  \label{sec: compute_complex}
This Subsection concerns the computational optimization of detection statistic calculations for transits of fixed duration but varying period and epoch. The specific parallel implementation of the detection pipeline is described in Appendix~\ref{ap: compute}.

The detection statistic (Equation~\ref{eq: detector}) includes the computation of a matrix-vector product $\Cov_{\mathbf{z}}^{-1} \mathbf{t}$ of order $O(N^2)$, where $N$ is the number of samples in the light curve. The inverse covariance calculations $\Cov_{\mathbf{z}}^{-1}$ are not a significant computational cost because they are independent of the transit signal $\bold{t}$ under study and are therefore calculated once per light curve. We optimize the computation of the term $\Cov_{\mathbf{z}}^{-1} \mathbf{t}$ over the transit search space $\mathbf{t} \in \mathbf{T}$ (Table \ref{tab:search_space}) by retaining intermediate results exploiting similarities for transit signals of fixed duration $d$; a similar approach is described by \citet{jenkins_2010}. A candidate transit signal $\mathbf{t}$ of duration $d$ may be decomposed into a number of disjoint single transit events $\mathbf{t}_{d, s}$ where $\Cov_{\mathbf{z}}^{-1} \mathbf{t} = \sum_{s} \Cov_{\mathbf{z}}^{-1} \mathbf{t}_{d, s}$. The matrix-vector product  for a single transit event $\Cov_{\mathbf{z}}^{-1} \mathbf{t}_{d, s}$ involves the summation of  $d$ matrix columns each of length $N$ which are contiguous in memory and can be accessed with an efficient cache hit rate; this specific computation consists of $d \cdot N$ additions.

For the transit template $\mathbf{t}_{d, s}$, we pre-compute $\Cov_{\mathbf{z}}^{-1} \mathbf{t}_{d, s} \in \mathcal{R}^{N}$ with different positional shifts of the transit template over step-sizes $\delta e$ in epoch. There are $\frac{N}{\delta e}$ possible positional shifts and each calculation requires $N d$ additions of the cached matrix $\Cov_{\mathbf{z}}^{-1}$. Therefore over all positional shifts a total of $\frac{N^2 d}{\delta e}$ additions are required. Over the full range of candidate durations $D$ these partial computations require $\sum_{d \in D} \frac{ N^2 d}{\delta e}$ additions.
These pre-computed matrix-vector products are held in intermediate memory.

The transit search space $\mathbf{T}$ can be decomposed over subspaces of fixed transit duration $\mathbf{T}_d$ (Appendix~\ref{ap: mu_t}). For each transit signal $\mathbf{t} \in \mathbf{T}_d$ we compute $\Cov_{\mathbf{z}}^{-1} \mathbf{t}$ by summing together the pre-computed single transit matrix-vector products.
If the average number of transit events per candidate transit $\mathbf{t} \in \mathbf{T}_d$ is $\mu_{\mathbf{t}}$ (Appendix~\ref{ap: mu_t}) then forming the sum of single-transit matrix-vector products to compute $\Cov_{\mathbf{z}}^{-1} \mathbf{t}$ for all $\mathbf{t} \in \mathbf{T}$ requires $|\mathbf{T}| N \mu_{\mathbf{t}}$ additions of the pre-computed vectors given prior reduction over $d$.

The final terms to be formed in the detection statistic are the inner product calculations $\mathbf{y}^T \Cov_{\mathbf{z}}^{-1} \mathbf{t}$ in the numerator and $\mathbf{t}^T \Cov_{\mathbf{z}}^{-1} \mathbf{t}$ in the denominator. Given the pre-computation of $\Cov_{\mathbf{z}}^{-1} \mathbf{t} \in \mathcal{R}^N$ described above, forming these additional inner products requires $N$ multiplications and additions respectively for a total $|\mathbf{T}| N$ operations over $\mathbf{t} \in \mathbf{T}$.

Combining all complexity terms, the transit search of $\mathbf{T}$ for a single light curve $\mathbf{y}$ is of complexity $O \left(|\mathbf{T}| N (1 + \mu_{\mathbf{t}} ) + \frac{\sum_{d \in D}d N^2}{\delta_e} \right)$. For the given transit search space, $|\mathbf{T}| \mu_{\mathbf{t}}$ scales as $N^2$ and $\mu_{\mathbf{t}}$ is constant (Appendix~\ref{ap: mu_t}). The leading-order complexity is therefore $O(|\mathbf{T}| N)$ or $O(N^3)$.

\subsubsection{High-performance-computing pipeline} \label{ap: compute}
SAP light curve data are provided per TESS observational sector in the MAST archive. Each sector file is 1.5 MB in size and a year of light curve data from the CVZ occupies $\sim$50 GB on hard disk. Detection statistics were computed per sector, as described in Section \ref{sec: compute_complex}, then combined across sectors (Equation~\ref{eq: final_det}) to compute detection tests for each light curve. We describe here the efficient implementation of these computations on the Blue Waters petascale system \citep{MENDES2015327}. We note that detection statistic computations are independent between light curves and between candidate transit durations.  

The inverse covariance $\Cov_{\mathbf{z}_i(q)}^{-1}$ of size  $N_q \times N_q$ is first computed once per sector $q$ for each light curve, where $N_q$ is the number of points in the light curve for sector $q$. This covariance matrix is $\sim$1.7 GB in size and is stored on hard disk. For each light curve, the covariances for all 13 sectors occupy 22 GB on disk. All covariances for all light curves in the CVZ for a single year require $\sim35$ TB of disk storage. 

A Blue Waters XE node consisted of 32 CPU cores and 32 GB of local memory.  Our unit of parallelism is a task, and 20 tasks were deployed on each XE node. A task is the computation of a set of detection tests for $\mathbf{t} \in \mathbf{T}_{d (q)}$, where $\mathbf{T}_{d (q)}$ is the realization of the transit search templates in $\mathbf{T}_d$ for sector $q$. This transit search subspace is of size $\approx 10^7$ and holds transit templates of fixed duration $d \in D$ for a single light curve sector $q$ of length $N_q$, over all period $P$ and epoch $e$ pairs in the sector for the parameter range defined in Table \ref{tab:search_space}, and further described in Appendix~\ref{ap: mu_t}. Tasks for the same light curve $i$ and sector $q$ were assigned to a the same node; there were $|D| = 10$ tasks per light curve. The common covariance matrix $\Cov_{\mathbf{z}_i(q)}^{-1}$ was shared as a memory-mapped object between all tasks on the node. Each task is highly parallel with no coupling or communication needed between tasks. In each task, single-transit event statistics of the form $\Cov_{\mathbf{z}_i(q)}^{-1} \mathbf{t}_{d, s}$ were first computed serially. Only the $N_q \cdot d$ covariance matrix elements for each single transit event $\mathbf{t}_{d, s}$ were read into local memory at a time. The memory required to hold all intermediate single-transit matrix-vector calculations $\Cov_{\mathbf{z}_i(q)}^{-1} \mathbf{t}_{d, s}$ per task was $\frac{N_q^2}{\delta e}$ (Appendix~\ref{sec: compute_complex}) 32-bit floating point values, of total size $\sim$ 0.4 GB. Within a single task, these single-transit calculations were combined to form detection tests in a serial manner for each period $P$ and epoch $e$ pair spanning $\mathbf{t} \in \mathbf{T}_{d (q)}$.

The output size of the detection tests of duration $d$ for a sector $q$ were $|\mathbf{T}_{d (q)}| \approx 10^7$ 32-bit floating point values, of total size 18 MB (therefore required $10^7$ writes per task).

The detection tests for a sector $q \in Q$ are intermediate as they are combined for all sectors to form composite detection statistics as described in Appendix \ref{sec: decompose}. The intermediate data output associated with a complete year of light curve data, for all candidate durations $d \in D$, is of size $|\mathbf{T}_{d (q)}| \cdot |D| \cdot |Q| \cdot |I|\sim$15 TB, where $|\mathbf{T}_{d (q)}|$ is the number of partial transits of duration $d$ and for a sector of data of length $N_q$, $|D| = 10$ is the number of searched candidate durations, $|Q| = 13$ is the total number of sectors spanning a year, and $|I| \sim 1500$ is the number of light curves per year of data. The detection test statistics, computed from this intermediate data, for all $\mathbf{t} \in \mathbf{T}$ for a single year-long light curve occupy 1 GB of total memory, and 5 TB for all light curves in the CVZ over 3 years.

\section{Loss in detection strength with a box transit approximation} \label{ap : box_approx}
Approaches to bound the detection strength due to general template mismatch for a form of matched filter which depends on template amplitude is provided by \citet[Section 4.2]{vantrees}. 
Here we quantify the exact mismatch between a true transit function $(\mathbf{t}_d \; : \; \|\mathbf{t}\| = 1)$ of transit duration $d$ and a box transit approximation $(\mathbf{b}_{d'} \; : \; \|\mathbf{b}_{d'}\| = 1)$ of duration $d' \leq d$ where we assume all signals are centered on $0$, for the form of matched filter used here which is invariant to template amplitude (shown in Appendix \ref{sec: decompose}) :

\begin{equation}
\mathbf{b}_{d'} [n] = 
\begin{cases}
    \frac{1}{\sqrt{d'}} & \text{if } |n|\leq \frac{d'}{2} \\
    0,              & \text{otherwise}
\end{cases}
\end{equation}
For mathematical convenience, we write the box transit normalized by duration as $\frac{1}{\sqrt{d'}}$, the detection statistics are invariant to this scaling. Consider the discretely sampled functions $\mathbf{t}_d$ and $\mathbf{b}_{d'}$ in vector form. The true transit signal $\mathbf{t}_d$ can be expanded as a linear combination along $\mathbf{b}_{d'}$ and an orthogonal vector $\mathbf{e}_d$, for which $\mathbf{b}_{d'}^T\mathbf{e}_d = \sum_{n} \mathbf{b}_{d'}[n]\mathbf{e}_d[n] = 0$, in the form:
\begin{equation}
    \mathbf{t}_d = (\mathbf{t}_d^T\mathbf{b}_{d'}) \mathbf{b}_{d'} + \mathbf{e}_d
\end{equation}
Trivially $\mathbf{e}_d = \mathbf{t}_d - \mu(\mathbf{t}_{d'})\mathbbm{1}_{d'}$ where $\mu$ is used to denote the sample average of $\mathbf{t}$ over $d'$: $\mu(\mathbf{t}_{d'}) = \frac{1}{d'}\sum_{n: |n| \leq \frac{d'}{2}} \mathbf{t}_{d'}[n]$ and $\mathbbm{1}_{d'}$ is a vector which is $1$ for $|n| \leq \frac{d'}{2}$ and $0$ for $|n| > \frac{d'}{2}$.
Omitting the covariance normalization for simplicity, the matched filter detector with template  $\mathbf{b}_{d'}$ is given by:
\begin{equation}
T(\mathbf{y} ; \mathbf{b}_{d'}) = \frac{\mathbf{y}^T\mathbf{b}_{d'}}{\sqrt{\mathbf{b}_{d'}^T \mathbf{b}_{d'} }} = \mathbf{y}^T\mathbf{b}_{d'}
\end{equation}
If $\mathbf{y} = A\mathbf{t}_d + \mathbf{n}$
\begin{equation}
\mathbb{E}[T(\mathbf{y} ; \mathbf{b}_{d'})] = \mathbb{E}[(A\mathbf{t}_d + \mathbf{n})^T\mathbf{b}_{d'}] = A\mathbf{t}_d^T\mathbf{b}_{d'} \\
= A ((\mathbf{t}_d^T\mathbf{b}_{d'}) \mathbf{b}_{d'} + \mathbf{e}_d)^T\mathbf{b}_d = A (\mathbf{t}_d^T\mathbf{b}_{d'})
\end{equation}
Repeating the same calculation with a matched filter using the true transit template (the Neyman-Pearson optimal detector):
\begin{equation}
\mathbb{E}[T(\mathbf{y} ; \mathbf{t}_d)] = A
\end{equation}
Therefore the loss in detection strength, using a box template of length $d'$ as compared to the true transit template of length $d$ is exactly given by $\mathbf{t}_d^T\mathbf{b}_{d'}$. 
This is the same quantity as the fitting factor in \citet{Allen} and \citet{Seader_2013} (Section 5) or $1 - \epsilon$ where $\epsilon$ is termed the template mismatch. 

In numerical simulations of limb darkened transit events over the range described in our simulated injection tests in Section \ref{sec: sim} and box functions of the same $d' = d$, this is approximately $\frac{\mathbb{E}[T(\mathbf{y} ; \mathbf{b}_d)]}{\mathbb{E}[T(\mathbf{y} ; \mathbf{t}_d)]} \sim 0.922$. When $d'$ is allowed to vary between discrete set of searched durations in Table \ref{tab:search_space}, the optimal $d'$ that minimizes the loss in detection strength attains $\frac{\mathbb{E}[T(\mathbf{y} ; \mathbf{b}_{d'})]}{\mathbb{E}[T(\mathbf{y} ; \mathbf{t}_d)]} \sim 0.960$. When $d'$ is allowed to take on any continuous value less than $d$, the optimal $d'$ attains $\frac{\mathbb{E}[T(\mathbf{y} ; \mathbf{b}_{d'})]}{\mathbb{E}[T(\mathbf{y} ; \mathbf{t}_d)]} \sim 0.968$. This suggests that on average a box transit approximation produces a minimal loss in detection power.

\citet{Seader_2013} also evaluate the fitting factor for a box template with Monte Carlo tests, however, in the context of the Kepler 30-minute cadence data. Their tests examine both mismatch due to shape with $d' = d$ and due to parameter discretization with $d' = \argmin_{d' \in D_{TPS}} \|d' - d\|$, where $D_{TPS}$ is a set of 14 logarithmic spaced durations between 1.5 to 15 hr. They note a loss in fitting factor of $0.0041$ (Table 1) for the latter case. In our tests where we vary $d'$, we instead minimize the loss in detection strength as $d' = \argmax_{d' \in D} \| \mathbf{t}_d^T \mathbf{b}_{d'}\|$ where $D$ is the discrete set of searched transit durations in Table \ref{tab:search_space}. As noted above, a $d' < d$ is optimal and for TESS 2-minute cadence data we observe an improvement in fitting factor of $0.038$ as compared to $d' = d$. 

\section{Transit search space properties} \label{ap: mu_t}
The transit template search space $\mathbf{T}$ is defined over the parameter range in transit duration $d$, period $P$, and epoch $e$ defined in Table~\ref{tab:search_space}. The search space $\mathbf{T}$ can be decomposed over subspace $\mathbf{T}_d$ for each discrete duration $d \in D$ as $\mathbf{T}=\bigcup_{d \in D} \mathbf{T}_d$.

Each duration subspace $\mathbf{T}_d$ spans a full sampling of period and epoch. The transit search templates are realized at a discrete sampling increment $\delta P$ in period and $\delta e$ in epoch. If a minimum of three transit events is required for detection of a transit signal in a light curve of length $N$, then the maximum detectable period is $P_{\rm{max}}=\frac{N}{2}$. Assuming a minimum period $\delta P$, the set of candidate periods is $P \in (\delta P) \cdot \{1,2,..,\frac{N}{2(\delta P)}\}$. For a given period $P$, the set of candidate epochs is $e \in (\delta e) \cdot \{1,2,...,\frac{P}{\delta e}\}$. 

The number of epochs for period $P$ is $N_e(P)=\frac{P}{\delta e}$. The total number of $(P,e)$ samples in each subspace $\mathbf{T}_d$, equivalently the subspace size $|\mathbf{T}_d|$, is obtained by summing over period:

\begin{align} \label{eq: search_size}
|\mathbf{T}_d| = \sum_{k =1}^{\frac{N}{2 (\delta P)}} N_e(k(\delta P)) = \frac{1}{2 (\delta e)} \left(\frac{N^2}{4 (\delta P)} + \frac{N}{2}\right)    
\end{align}

The size of the full transit search space is,
\begin{equation}
    |\mathbf{T}| = |D|\cdot |\mathbf{T}_d|= \frac{|D|}{2 (\delta e)}\left(\frac{N^2}{4 (\delta P)} + \frac{N}{2}\right) \approx \frac{|D| N^2 }{8 (\delta e) (\delta P)}
\end{equation}

The number of transit events for fixed period $P$ in a light curve of length $N$ is $\frac{N}{P}$. The total number of transit events $N_{t,d}$ in each transit search subspace $\mathbf{T}_d$ is therefore,

\begin{align}
    N_{t,d} = \sum_{k=1}^{\frac{N}{2 (\delta P)}} \frac{N_e(k(\delta P)) N}{k(\delta P)} =  \frac{N^2}{2 (\delta e) (\delta  P)}
\end{align}
The average number of transit events $\mu_t$ per template in $\mathbf{T}_d$ is constant:
\begin{align}
    \mu_\mathbf{t} = \frac{N_{t,d}}{|\mathbf{T}_d|} \sim 4
\end{align} 

From Table~\ref{tab:search_space}, $|D| = 10$ and $ \delta P = \delta e = 5 \triangle_{LC}$ 2-min cadence samples (or 10 min): then $|\mathbf{T}| \sim \frac{N^2}{20}$. For $N \sim 2 \cdot 10^5$ (a year of 2-minute cadence samples) this yields $|\mathbf{T}| \leq 10^9$. In practice, $|\mathbf{T}| \approx 10^8$ as transits that occur on masked data are excluded, and a larger step size of 20 min is used for candidate periods greater than 27 days (Table~\ref{tab:search_space}). 

\bibliography{mybib}
\bibliographystyle{aasjournal}

\end{document}